\documentclass[11pt]{article}
\usepackage[margin=1in]{geometry}
\usepackage{chihao}
\usepackage{pgfplots}
\pgfplotsset{compat=1.18}
\usepackage{subfigure}
\usepackage{multirow}
\usepackage{comment}
\usepackage{verbatim}

\newcommand{\red}[1]{{\color{red}{#1}}}
\newcommand{\blue}[1]{{\color{blue}{#1}}}

\newcommand{\UniformPricing}{\textsf{Uniform Pricing}}

\renewcommand{\iff}{\Leftrightarrow}
\renewcommand{\implies}{\Rightarrow}

\newcommand{\ignore}[1]{}
\newcommand{\DKL}{\!{KL}}

\newcommand{\tTheta}{\wt{\Theta}}

\newcommand{\bv}{\boldsymbol{v}}

\newcommand{\bF}{\boldsymbol{F}}

\newcommand{\bH}{\boldsymbol{H}}

\newcommand{\bX}{\boldsymbol{X}}

\title{The Query Complexity of Uniform Pricing}

\author{
Houshuang Chen\thanks{Shanghai Jiao Tong University. Email: {\tt chenhoushuang@sjtu.edu.cn}}
\and
Yaonan Jin\thanks{Hong Kong University of Science and Technology. Email: {\tt jinyaonan1996@gmail.com}}
\and
Pinyan Lu\thanks{Shanghai University of Finance and Economics, Laboratory of Interdisciplinary Research of Computation and Economics (SUFE), \& Huawei. Email: {\tt lu.pinyan@mail.shufe.edu.cn}}
\and
Chihao Zhang\thanks{Shanghai Jiao Tong University. Email: {\tt chihao@sjtu.edu.cn}}
}
\date{}

\begin{document}

\maketitle

\thispagestyle{empty}

\begin{abstract}
Real-world pricing mechanisms are typically optimized using training data, a setting corresponding to the \textit{pricing query complexity} problem in Mechanism Design.
The previous work \cite{LSTW23soda} studies the \textit{single-distribution} case,\footnote{\label{footnote:tilde}The $\widetilde{\Theta}$ notation omits polylogarithmic factors.} with tight bounds of $\widetilde{\Theta}(\varepsilon^{-3})$ for a \textit{general} distribution and $\widetilde{\Theta}(\varepsilon^{-2})$ for either a \textit{regular} or \textit{monotone-hazard-rate (MHR)} distribution, where $\varepsilon \in (0, 1)$ denotes the (additive) revenue loss of a learned uniform price relative to the Bayesian-optimal uniform price.

This can be directly interpreted as ``the query complexity of the {\em \textsf{Uniform Pricing}} mechanism, in the \textit{single-distribution} case''.
Yet in the \textit{multi-distribution} case, can the regularity and MHR conditions still lead to improvements over the tight bound $\widetilde{\Theta}(\varepsilon^{-3})$ for general distributions?
We answer this question in the negative, by establishing a (near-)matching lower bound $\Omega(\varepsilon^{-3})$ for either \textit{two regular distributions} or \textit{three MHR distributions}.

We also address the \textit{regret minimization} problem and, in comparison with the folklore upper bound $\widetilde{O}(T^{2 / 3})$ for general distributions (see, e.g., \cite{SW24}), establish a (near-)matching lower bound $\Omega(T^{2 / 3})$ for either \textit{two regular distributions} or \textit{three MHR distributions}, via a black-box reduction. Again, this is in stark contrast to the tight bound $\widetilde{\Theta}(T^{1 / 2})$ for a single regular or MHR distribution.

\end{abstract}

\newpage

\setcounter{page}{1}

\section{Introduction}
\label{sec:intro}

{\UniformPricing} serves as a foundational mechanism in both economic theory \cite{ACPRW14} and real-world markets \cite{DG19}. It describes a scenario where a platform sets a uniform price for $n \ge 1$ buyers and the transaction succeeds when at least one buyer is willing to accept this uniform price. From e-commerce platforms pricing products to cloud providers setting subscription fees, {\UniformPricing}'s ubiquity and operational simplicity belie a crucial challenge, especially in repeated interactions:
\begin{quote}
    \textit{How to learn --- information-efficiently --- an (approximately) revenue-optimal uniform price?}
\end{quote}
The arguably ``most reasonable'' information revealed in (a trial of) {\UniformPricing} is the success or failure of the transaction.\footnote{This is mainly because {\UniformPricing} is a pricing mechanism, where buyers make take-it-or-leave-it decisions. In contrast, for a (truthful) auction mechanism, the arguably ``most reasonable'' information is samples of buyers' values. (For a survey of the ``sample complexity of mechanism design'' literature, the interested reader can refer to the early works \cite{DRY15,CR14,HMR18} as well as the recent work \cite{FJ25} and the references therein.)} Accordingly, a platform can learn and optimize {\UniformPricing} using \textit{pricing queries} \cite{LSTW23soda, LSTW23ec, SW24, TW25}, namely trials of this mechanism itself. The efficiency of this learning process is measured by two canonical metrics (see \Cref{sec:prelim,sec:regret} for their formal definitions):
\begin{itemize}
    \item \textit{Query Complexity:}
    How many trials does a (possibly adaptive) pricing strategy require to output an $\eps \in (0, 1)$-approximately revenue-optimal uniform price.
    
    \item \textit{Minimax Regret:}
    Over $T \ge 1$ trials, compared with an (exactly) revenue-optimal uniform price, how much a (possibly adaptive) pricing strategy will lose in cumulative revenue.
\end{itemize}

\subsection{Previous Works: The Single-Distribution Case}

In the \textit{single-distribution} case ($n = 1$), the works \cite{LSTW23soda,SW24} have obtained a clear set of conclusions:
\begin{quote}
    \textit{Good single-distribution structure (regularity/MHR) can greatly improve the learning efficiency.}
\end{quote}
(i)~For a single \textit{general} distribution (without distributional structure), the query complexity scales as $\tTheta(\eps^{-3})$, and the minimax regret scales as $\tTheta(T^{2/3})$.\textsuperscript{\ref{footnote:tilde}}\\
(ii)~For a single \textit{regular} distribution \cite{M81} or a single \textit{monotone hazard rate (MHR)} distribution \cite{BMP63}\footnote{A distribution $F$ satisfies the (relatively weaker) regularity condition \cite{M81} when its \textit{virtual value function} $\phi(v) = v - \frac{1 - F(v)}{f(v)}$ is nondecreasing, and satisfies the (relatively stronger) MHR condition \cite{BMP63} when its \textit{hazard rate function} $h(v) = \frac{f(v)}{1 - F(v)}$ is nondecreasing; see \Cref{sec:prelim} for more details.} --- two standard distributional conditions in the literature --- the query complexity improves to $\tTheta(\eps^{-2})$, and the minimax regret improves to $\tTheta(T^{1 / 2})$. Basically, the regularity/MHR condition imposes ``convexity-like properties''\footnote{\label{footnote:convexity}More precisely, for a \textit{regular} distribution $F$, (in a parametric equation form) the \textit{revenue-quantile curve} $(1 - F(p), p \cdot (1 - F(p)))$ is \textit{concave}. And for a \textit{MHR} distribution $F$, the \textit{cumulative hazard rate function} $H(p) = -\ln(1 - F(p))$ is \textit{convex}.} and, thus, enables ``binary-like search'' for an (approximately) revenue-optimal price.

\subsection{This Work: The Multi-Distribution Case}

The above results can be seamlessly interpreted as the learning efficiency of {\UniformPricing}, in the \textit{single-distribution} case ($n = 1$). Instead, this work addresses the extension to the \textit{multi-distribution} case ($n \ge 2$), i.e., a scenario with $n \geq 2$ (independent) distributions drawn from a common distribution class --- general, regular, or MHR. Of particular interest is the following question:
\begin{quote}
    \textit{Can good multi-distribution structure (regularity/MHR) still improve the learning efficiency?}
\end{quote}

The multi-distribution case ($n \ge 2$), apart from being a seamless generalization, is worth investigating in various additional aspects.
Firstly, value distributions in real-world markets can rarely be of a single type. For example, ride-hailing platforms such as Uber and Lyft have multi-type user values; time-sensitive riders (e.g., for work) might prioritize speed over cost, while off-peak riders often prefer cheaper options.
Also, online shopping platforms such as Amazon often tier prices (even on the same product) for different types of buyers; urgent buyers (needing next-day deliveries) must pay full price, bulk shoppers may get volume discounts, and deal-seekers can wait for flash sales.


In addition, from a technical perspective, the success or failure of (the transaction in) {\UniformPricing} depends on the \textit{highest value} $\max_{i \in [n]} v_{i}$ (across all distributions $v_{i} \sim F_{i}$ for $i \in [n]$) and the corresponding \textit{first-order distribution} $F(p) = \prod_{i \in [n]} F_{i}(p)$. In this vein:\\
(i)~For multiple \textit{general} distributions, their first-order distribution $F(p)$ possesses no specific structure, so the tight query complexity $\tTheta(\eps^{-3})$ and the tight minimax regret $\tTheta(T^{2/3})$ are the same as before.\\
(ii)~For multiple \textit{regular/MHR} distributions, their first-order distribution $F(p)$ may violate the regularity/ MHR condition but, to a certain extent, still possesses good structure.\footnote{For example, for multiple \textit{regular/MHR} distributions, their first-order distribution $F(p)$ satisfies the \textit{quasi-regular/quasi-MHR} condition --- a natural relaxation/generalization of regularity/MHR introduced by the recent work \cite{FJ25}.

A distribution $F$ satisfies the (relatively weaker) quasi-regularity condition \cite{FJ25} when its \textit{``conditional expected''} virtual value function $\phi_{CE}(v) = \mathbb{E}[\phi(x) | x \leq v]$ is nondecreasing, and satisfies the (relatively stronger) quasi-MHR condition \cite{FJ25} when its \textit{``normalized cumulative''} hazard rate function $h_{NC}(v) = \frac{1}{v} \int_{0}^{v} h(x) \dd x$ is nondecreasing. So, the quasi-regularity (resp.\ quasi-MHR) condition relaxes the \textit{pointwise monotonicity} of a virtual value function (resp.\ a hazard rate function) to \textit{on-average monotonicity}.}
Hence, it is important to determine whether such ``moderate'' distributional structures can still improve the learning efficiency.

\subsection{Our Contributions}

We answer the above questions on the learning efficiency of {\UniformPricing}, showing a sharp dichotomy between the single-distribution case ($n = 1$) and the multi-distribution case ($n \ge 2$):
\begin{quote}
    \textit{Good multi-distribution structure (regularity/MHR) \textbf{cannot} improve the learning efficiency.}
\end{quote}
Namely, in the multi-distribution case ($n \ge 2$), both the query complexity and the minimax regret for \textit{regular/MHR} distributions (essentially) revert to those for \textit{general} distributions.
In more detail, our conceptual and technical contributions can be divided into the following three categories.

\vspace{.1in}
\noindent
\textbf{1. Matching Lower Bounds for Multiple Well-Structured Distributions.}
We establish the following hardness results for both metrics --- query complexity and minimax regret --- showing that good distributional structure provides (almost) no benefit in the multi-distribution case ($n \ge 2$):
\begin{itemize}
    \item For $n \geq 2$ regular distributions, the regularity condition cannot help in learning {\UniformPricing}:\\
    We prove a query complexity lower bound of $\Omega(\eps^{-3})$ and a minimax regret lower bound of $\Omega(T^{2/3})$, matching the tight bounds $\tTheta(\eps^{-3})$ and $\tTheta(T^{2/3})$ for general distributions.
    
    \item For $n \geq 3$ MHR distributions, even the stronger MHR condition cannot help learn {\UniformPricing}:\\
    Again, we establish a matching query complexity lower bound of $\Omega(\eps^{-3})$ and a matching minimax regret lower bound of $\Omega(T^{2/3})$.
    
    \item For $n = 2$ MHR distributions, we prove a query complexity lower bound of $\Omega(\eps^{-5/2})$ and a minimax regret lower bound of $\Omega(T^{3/5})$, leaving small gaps relative to the tight bounds $\tTheta(\eps^{-3})$ and $\tTheta(T^{2/3})$ for general/regular distributions.
\end{itemize}
These hardness results contrast sharply with the single-distribution case ($n = 1$) \cite{LSTW23soda,SW24}, where regularity/MHR significantly improves the learning efficiency (such as reducing the query complexity from $\tTheta(\eps^{-3})$ to $\tTheta(\eps^{-2})$).
So, in more competitive scenarios, such as ride-hailing and online shopping with $n \ge 2$ distributions, a platform must be more careful about the design of its pricing strategies.

\vspace{.1in}
\noindent
\textbf{2. Insights behind Lower-Bound Construction.}
As noted, the learning of a revenue-optimal uniform price $p^{\sf opt}$ (say) relies on the underlying \textit{first-order distribution} $F(p) = \prod_{i \in [n]} F_{i}(p)$ and the corresponding \textit{revenue function} $R(p) = p \cdot (1 - F(p))$.

For a single regular/MHR distribution, the revenue function $R(p)$ turns out to exhibit ``convexity-like properties'' \cite{SW24},\textsuperscript{\ref{footnote:convexity}} which then enables ``binary-like search'' of $p^{\sf opt}$ (or its good enough approximations). This accounts for the improvements to $\tTheta(\eps^{-2})$ and $\tTheta(T^{1/2})$ (over the tight bounds $\tTheta(\eps^{-3})$ and $\tTheta(T^{2/3})$ for general distributions).


For multiple regular/MHR distributions, to establish our hardness results (that match the tight bounds for general distributions), our lower-bound construction must break the above ``convexity-like properties''.
We achieve this by leveraging the competition across individual regular/MHR distributions. Namely, even if all distributions $F_{i}$ for $i \in [n]$ are regular/MHR, the revenue function $R(p)$ can have two features:
\begin{itemize}
    \item \textit{Global Flatness:}
    $R(p)$ varies by at most $O(\eps)$, over a wide enough region $I$ promised to contain $p^{\sf opt}$.
    
    \item \textit{Local Sharpness:}
    $R(p)$ jumps up by at least $\Omega(\eps)$ on narrow enough sub-intervals of $I$, e.g., at $p^{\sf opt}$.
\end{itemize}
It turns out that the global flatness means ``the verification of $p^{\sf opt}$'s revenue-optimality'' is inefficient, and the local sharpness means ``the search for (a narrow enough sub-interval that contains) $p^{\sf opt}$'' is inefficient. In combination, this lower-bound construction scheme is sufficient to establish our hardness results.

\vspace{.1in}
\noindent
\textbf{3. A Unified Framework for Lower-Bound Analysis.}
To obtain our hardness results, we have further developed a unified framework for lower-bound analysis, adapting it to the specific contexts considered. Specifically, this framework consists of four components:
\begin{itemize}
    \item \textit{Base Instance:}
    Construct a suitable base instance $\bF^{*} = \otimes_{i \in [n]} F_{i}^{*}$, such that the revenue function $R^{*}$ is exactly flat (i.e., all uniform prices $p \in I$ are equally revenue-optimal) over a wide enough region $I$.
    
    \item \textit{Hard Instances:}
    Each hard instance $\bF^{k}$ for $k = 1, 2, \dots, K$ (say $K = \Omega(\eps^{-1})$) modifies the base instance $\bF^{*}$ on some sub-interval $I^{k} \subseteq I$, in such a way:\\
    (i)~Each hard instance $\bF^{k}$ retains the same distributional structure (regularity/MHR) as $\bF^{*}$.\\
    (ii)~$R^{k}(p) > R^{*}(p) + \eps$ for some $p \in I^{k}$, some modified revenue can exceed the base by more than $\eps$.\\
    (iii)~Different modification sub-intervals $I^{k}$ for $k = 1, 2, \dots, K$ are disjoint.
    
    \item \textit{A Reduction from Query Complexity to Instance-Identification:}
    To output an $\eps$-approximately revenue-optimal uniform price (say) in each possibility $k \in [K]$, a pricing strategy must identify the actual modification sub-interval $I^{k}$ and the actual hard instance $\bF^{k}$.
    Also, information-theoretic arguments show that a single hard instance $\bF^{k}$ requires $\Omega(\eps^{-2})$ queries to identify. Thus, a combination of both arguments gives a query complexity lower bound of $K \cdot \Omega(\eps^{-2}) = \Omega(\eps^{-3})$ (say).
    
    \item \textit{A Reduction from Minimax Regret to Instance-Identification:}
    This reduction simply adapts the above one, from the query complexity problem to the minimax regret problem; see \Cref{sec:regret} for details.
\end{itemize}
We believe that our unified framework is general-purpose --- {\UniformPricing} as a canonical mechanism is a representative instantiation --- and can find far more applications in future research.

\vspace{.1in}
\noindent
\textbf{Organization.}
For ease of presentation, we focus on the query complexity problem throughout \Cref{sec:prelim,sec:regular,sec:three-MHR,sec:two-MHR}.
In \Cref{sec:prelim}, we provide preliminaries.
In \Cref{sec:regular}, we prove an $\Omega(\eps^{-3})$ lower bound for $n \geq 2$ regular distributions.
In \Cref{sec:three-MHR}, we prove an $\Omega(\eps^{-3})$ lower bound for $n \geq 3$ MHR distributions.
In \Cref{sec:two-MHR}, we prove an $\Omega(\eps^{-5 / 2})$ lower bound for $n = 2$ MHR distributions.
Finally, in \Cref{sec:regret}, we extend these query complexity lower bounds to the corresponding minimax regret lower bounds.

\section{Preliminaries}
\label{sec:prelim}

For a positive integer $n \ge 1$, we denote $[n] \defeq \{1, 2, \dots, n\}$.

\subsection{Probability and Distribution}

Consider a probability space $(\Omega, \+F, \bb{P})$. For a sequence of random variables $\bX = (X_{i})_{i \in [n]}$, we denote by $\bb{P}_{\bX}(A)$ the \textit{pushforward measure} of $\bb{P}$ by these random variables $\bX$, for every measurable set $A$:\footnote{For readers unfamiliar with this notion, when $\bX$ consists of a single random variable $X_{1}$, we can informally interpret $\bb{P}_{\bX}$ as the marginal probability of $X_{1}$ when $X_{1}$ is discrete, or the marginal density of $X_{1}$ when $X_{1}$ is absolutely continuous with respect to the Lebesgue measure. We adopt this more general measure-theoretic notion since the random variables $\bX = (X_{i})_{i \in [n]}$ might be neither discrete nor absolutely continuous.}
\begin{align*}
    \bb{P}_{\bX}(A) ~\defeq~ \bb{P}[\{\omega \in \Omega: \bX \in A\}].
\end{align*}
For a sub $\sigma$-algebra $\+F' \subseteq \+F$, we also denote by $\bb{P}_{\bX | \+F'}(A)$ the \textit{conditional pushforward measure} given $\+F'$:
\begin{align*}
    \bb{P}_{\bX | \+F'}(A) ~\defeq~ \bb{P}[\{\omega \in \Omega: \bX \in A\}\,|\,\+F'].
\end{align*}

Regarding a \textit{single-dimensional distribution} $F$, without ambiguity, we abuse the notation $F$ also for its \textit{cumulative distribution function (CDF)}.
Throughout this paper, we consider left-continuous CDF's, namely $F(p) \defeq \bb{P}_{v \sim F}[v < p]$ for $p \in [-\infty, +\infty]$; we prefer this shift from convention, since a buyer with a random value $v \sim F$ is willing to buy a price-$p$ item with probability $\bb{P}[v \ge p]$, rather than $\bb{P}[v > p]$.

The following \Cref{def:regular,def:MHR} introduces two canonical distribution families, the family of \textit{regular} distributions \cite{M81} and the family of \textit{monotone-hazard-rate (MHR)} distributions \cite{BMP63} --- a regular or MHR distribution $F$ always has a well-defined (generalized) \textit{probability density function (PDF)} $f$.

\begin{definition}[Regular Distributions \cite{M81}]
\label{def:regular}
A distribution $F$ satisfies the \textit{regularity} condition when its virtual value function $\phi(x) \defeq x - \frac{1 - F(x)}{f(x)}$ is \textit{nondecreasing} over its support.
\end{definition}

\begin{definition}[MHR Distributions \cite{BMP63}]
\label{def:MHR}
A distribution $F$ satisfies the \textit{monotone-hazard-rate (MHR)} condition when its hazard rate function $h(x) \defeq \frac{f(x)}{1 - F(x)}$ is \textit{nondecreasing} over its support.
\end{definition}

\noindent
There are other alternative/equivalent definitions of these two conditions; for details, the interested reader can reference the textbook \cite{H13} and the recent work \cite{FJ25}.
By definitions, a MHR distribution must be a regular distribution, but the converse is incorrect in general; distributions like $F(x) = \max\tp{\frac{x - 1}{x}, 0}$ for $x \in [0, +\infty]$ are regular but non-MHR.

In addition, the following \Cref{cla:Jensen} presents \textit{Jensen's Inequality} in the context of probability theory.

\begin{claim}[{Jensen's Inequality \cite[Theorem~2.6.2]{CT06}}]
\label{cla:Jensen}
If $X$ is a random variable and $g$ is a convex function, then $g(\mathbb{E}[X]) \le \mathbb{E}[g(X)]$.
\end{claim}

\subsection{Uniform Pricing}

In single-item mechanism design, a seller aims to sell an indivisible item to $n \ge 1$ buyers with \textit{independent} value distributions $\bF = \otimes_{i = 1}^{n} F_{i}$.
Specifically, the {\UniformPricing} mechanism posts a uniform price $p \ge 0$ on the item and sells it to any buyer (such as the first coming one) willing to pay this price; this results in the \textit{first-order value distribution} $F$ and the \textit{revenue function} $R$.
\begin{align*}
    F(p)
    &\textstyle ~\defeq~ \bb{P}_{\bv \sim \bF}[(\max_{i \in [n]} v_{i}) < p]
    ~=~ \prod_{i = 1}^{n} F_{i}(p),
    &&\textstyle \forall p \ge 0,\\
    R(p)
    &\textstyle ~\defeq~ p \cdot \bb{P}_{\bv \sim \bF}[(\max_{i \in [n]} v_{i}) \ge p]
    ~=~ p \cdot \big(1 - F(p)\big),
    &&\textstyle \forall p \ge 0.
\end{align*}

In the bulk of this paper, we will study (a generalized version of) the \textit{pricing query complexity} problem; to make the problem interesting, we follow the previous works \cite{LSTW23soda, LSTW23ec, TW25} and consider $[0, 1]$-supported value distributions. (Thus, the \textit{optimal uniform price} $p^{\sf opt} = p^{\sf opt}(\bF) \defeq \argmax_{p \in [0,1]} R(p)$ is well-defined and lies in the support $[0, 1]$.)
A pricing algorithm $\+A$ works as follows:
\begin{itemize}
    \item At the beginning, $\+A$ has no information of the value distributions $\bF$ (except for their independence and $[0, 1]$ support).
    
    \item $\+A$ acquires information of the value distributions $\bF$ through \textit{pricing queries}.\footnote{In this way, pricing algorithms come in two flavors, \textit{adaptive} and \textit{non-adaptive}: an adaptive one can determine a query price $p^{t}$ based on the information acquired thus far, while a non-adaptive one must determine all query prices $p^{1}, p^{2}, \dots$ in advance.} Each time $t = 1, 2, \dots$, $\+A$ posts a \textit{query price} $p^{t}$ and acquires \textit{binary feedback} \allowdisplaybreaks $z^{t} = z^{t}(p^{t}) \defeq \bb{1}[(\max_{i \in [n]} v_{i}^{t}) \ge p^{t}] \in \{0, 1\}$ based on an independent draw $\bv^{t} \sim \bF$; this gives $\bb{P}[z^{t} = 0] = F(p^{t})$ and $\bb{P}[z^{t} = 1] = 1 - F(p^{t})$, i.e., an independent trial of the {\UniformPricing} mechanism --- whether the sale using a uniform price $p^{t}$ succeeds or not.
    
    \item At the termination, $\+A$ needs to output a price $p^{\+A}$.
\end{itemize}
The pricing query complexity problem asks, over the randomness of both the value distributions $\bF$ and the pricing algorithm $\+A$ itself, \textit{how many pricing queries are sufficient and necessary to succeed in outputting a ``good enough'' price $p^{\+A}$}: given $\eps \in (0, 1)$,
\begin{align*}
    \textstyle
    \bb{P}_{\bF, \+A}\big[R(p^{\+A}) \ge R(p^{\sf opt}) - \eps\big] ~\ge~ \frac{2}{3}.
\end{align*}
Here, the $\frac{2}{3}$ success probability is a standard convention in designing probabilistic algorithms (and can be replaced by any other constant strictly larger than $\frac{1}{2}$).

Later in \Cref{sec:regret}, we will study another related problem, the \textit{regret minimization} problem.

\subsection{Information Theory}

Let $\bb{P}^{*}$ and $\bb{P}$ be two probability measures on the same measurable space $(\Omega, \+F)$. When $\bb{P}^{*}$ is absolutely continuous with respect to $\bb{P}$, their \textit{Kullback-Leibler (KL) divergence} \cite[Chapter~2.3]{CT06} is given by
\begin{align*}
    \textstyle
    \DKL(\bb{P}^{*}, \bb{P})
    ~\defeq~ \bb{E}_{\bb{P}^{*}}\Big[\ln(\dv{\bb{P}^{*}}{\bb{P}})\Big],
\end{align*}
where $\dv{\bb{P}^{*}}{\bb{P}}$ is the Radon-Nikodym derivative. Without ambiguity, we abuse the notation $\DKL(p, q)$ to denote the KL divergence between two Bernoulli distributions with parameters $p, q \in [0, 1]$:
\begin{align*}
    \textstyle
    \DKL(p, q)
    ~\defeq~ p \ln(\frac{p}{q}) + (1 - p) \ln(\frac{1 - p}{1 - q}).
\end{align*}

The following \Cref{cla:DKL-convex,cla:DKL-UB,lem:kl-bound} will be useful in our later proofs.

\begin{claim}[Convexity of $\DKL(p, q)$]
\label{cla:DKL-convex}
The function $\DKL(p, q)$ is convex on $(p, q) \in [0, 1]^{2}$.
\end{claim}

\begin{proof}
By elementary algebra, the function $\DKL(p, q)$'s Hessian matrix $\bH_{\DKL}$ is given by
\begin{align*}
    \bH_{\DKL}
    ~\defeq~
    \begin{bmatrix}
	    \frac{\partial^{2} \DKL}{\partial p^{2}} & \frac{\partial^{2} \DKL}{\partial p \partial q}\\
	    \frac{\partial^{2} \DKL}{\partial q \partial p} & \frac{\partial^{2} \DKL}{\partial q^{2}}
    \end{bmatrix}
    ~=~
    \begin{bmatrix}
	    \frac{1}{p} + \frac{1}{1 - p} & -\frac{1}{q} - \frac{1}{1 - q}\\
	    -\frac{1}{q} - \frac{1}{1 - q} & \frac{p}{q^{2}} + \frac{1 - p}{(1 - q)^{2}}
    \end{bmatrix}.
\end{align*}
Since $\bH_{\DKL}$ has \textit{nonnegative} diagonal elements as well as a \textit{nonnegative} determinant
\begin{align*}
    \textstyle
    |\bH_{\DKL}|
    &\textstyle ~=~ \big(\frac{1}{p}+\frac{1}{1 - p}\big) \cdot \big(\frac{p}{q^{2}} + \frac{1 - p}{(1 - q)^{2}}\big) - \big(\frac{1}{q} + \frac{1}{1 - q}\big)^{2}\\
	&\textstyle ~=~ \frac{1 - p}{p \cdot (1 - q)^{2}} + \frac{p}{(1 - p) \cdot q^{2}} - \frac{2}{q \cdot (1 - q)}\\
	&\textstyle ~\ge~ 0,
\end{align*}
the function $\DKL(p, q)$ is \textit{convex} on $(p, q) \in [0, 1]^{2}$. This finishes the proof of \Cref{cla:DKL-convex}.
\end{proof}

\begin{claim}[Upper Bounds of $\DKL(p, q)$]
\label{cla:DKL-UB}
$\DKL(p, q) \le 3 \cdot (p - q)^{2}$, for $p \in [\frac{1}{7}, \frac{6}{7}]$ and $q \in [p - \frac{1}{12}, p]$.
\end{claim}

\begin{proof}
For notational brevity, let $\delta \defeq p - q \in [0, \frac{1}{12}]$; note that $\frac{\delta}{p}, \frac{\delta}{1 - p} \in [0, \frac{7}{12}]$. We deduce that
\begin{align*}
    \DKL(p, q)
    &\textstyle ~=~
    -p \ln(1 - \frac{\delta}{p}) - (1 - p) \ln(1 + \frac{\delta}{1 - p})\\
    &\textstyle ~\le~ p \cdot \Big(\frac{\delta}{p} + \frac{\delta^{2}}{p^{2}}\Big) - (1 - p) \cdot \Big(\frac{\delta}{1 - p} - \frac{1}{2} \cdot \frac{\delta^{2}}{(1 - p)^{2}}\Big)\\
    &\textstyle ~=~ \frac{\delta^{2}}{p} + \frac{\delta^{2}}{2 \cdot (1 - p)}\\
    &\textstyle ~\le~ 3\delta^{2}.
\end{align*}
Here the second step uses $-\ln(1 - x) \le x + x^{2}$ and $\ln(1 + x) \ge x - \frac{1}{2}x^{2}$, for $x \in [0, \frac{7}{12}]$. And the last step uses $\frac{1}{p} + \frac{1}{2 \cdot (1 - p)} \le \frac{3}{2} + \sqrt{2} \approx 2.9142$. This finishes the proof of \Cref{cla:DKL-UB}.
\end{proof}

\begin{claim}[Pricing Algorithms]
\label{lem:kl-bound}
Consider two instances $\bF^{*}$ and $\bF$ that their first-order CDF's are identical, $F(p) = F^{*}(p)$ for $p \notin P$, everywhere except for a measurable subset $P \subseteq [0, 1]$.

Let $\bb{P}^{*}$ and $\bb{P}$ be the probability measures, on the same measurable space $(\Omega, \+F)$, by running a pricing algorithm $\+A$ respectively on $\bF^{*}$ and $\bF$.
Then, for any random event $\@E \in \+F$,
\begin{align*}
    \textstyle
    \DKL\big(\bb{P}^{*}[\@E],\, \bb{P}[\@E]\big)
    ~\le~ \bb{E}_{\bb{P}^{*}}[T^{P}] \cdot \big(\max_{p \in P} \DKL(F^{*}(p),\, F(p))\big).
\end{align*}
where $T^{P} \defeq |\{t \in [T]:\, p^{t} \in P\}|$ denotes how many pricing queries are made using query prices in $P$.
\end{claim}

\begin{proof}
Since the conclusion trivially holds when $\mathbb{E}[T^{P}] = +\infty$, below we focus on the case $\mathbb{E}[T^{P}] < +\infty$. By the \textit{data-processing inequality} for KL divergence \cite[Chapter~2.8]{CT06},
\begin{align*}
    \DKL\big(\bb{P}^{*}[\@E],\, \bb{P}[\@E]\big)
    ~\le~ \DKL(\bb{P}^{*},\, \bb{P}).
\end{align*}
We denote by $\@H^{t} \defeq \{(p^{\tau}, z^{\tau})\}_{\tau = 1}^{t - 1}$ the query prices posted and the binary feedback acquired before time $t$; let $\@H^{1} \defeq \emptyset$ for notational consistency. We can deduce that
\begin{align*}
    \DKL\big(\bb{P}^{*},\, \bb{P}\big)
    &\textstyle ~=~ \bb{E}_{\bb{P}^{*}}\bigg[\sum_{t \in [T]} \bigg(\ln(\dv{\bb{P}^{*}_{z^{t} | \@H^{t} \cup \{p^{t}\}}}{\bb{P}_{z^{t} | \@H^{t} \cup \{p^{t}\}}}) + \ln(\dv{\bb{P}^{*}_{p^{t} | \@H^{t}}}{\bb{P}_{p^{t} | \@H^{t}}})\bigg)\bigg]\\
    &\textstyle ~=~ \bb{E}_{\bb{P}^{*}}\bigg[\sum_{t \in [T]} \ln(\dv{\bb{P}^{*}_{z^{t} | p^{t}}}{\bb{P}_{z^{t} | p^{t}}})\bigg]\\
    &\textstyle ~=~ \E[\bb P^*]{\sum_{t \in [T]:p^t\in P} \ln(\dv{\bb{P}^{*}_{z^{t} | p^{t}}}{\bb{P}_{z^{t} | p^{t}}})}\\
    &\textstyle ~=~ \bb{E}_{\bb{P}^{*}}\bigg[\sum_{t \in [T]:\, p^{t} \in P} \bb{E}_{\bb{P}^{*}_{z^{t} | p^{t}}}\bigg[\ln(\dv{\bb{P}^{*}_{z^{t} | p^{t}}}{\bb{P}_{z^{t} | p^{t}}})\bigg]\bigg]\\
    &\textstyle ~\le~ \bb{E}_{\bb{P}^{*}}\bigg[\sum_{t \in [T]:\, p^{t} \in P} \big(\max_{p \in P} \DKL(F^{*}(p),\, F(p))\big)\bigg]\\
    &\textstyle ~=~ \bb{E}_{\bb{P}^{*}}[T^{P}] \cdot \big(\max_{p \in P} \DKL(F^{*}(p),\, F(p))\big),
\end{align*}
Here  the third step holds since $F(p) = F^{*}(p)$ for $p \notin P$.
And the last step uses \textit{Wald's equation} \cite{W44}.

This finishes the proof of \Cref{lem:kl-bound}.
\end{proof}

\section{\texorpdfstring{$\Omega(\eps^{-3})$ Lower Bound for Two Regular Distributions}{}}
\label{sec:regular}

In this section, we investigate the query complexity of {\UniformPricing} in the setting with \textit{regular} distributions. Specifically, we will establish the following \Cref{thm:two-regular}.

\begin{theorem}
\label{thm:two-regular}
For two (or more) regular distributions, the query complexity of {\UniformPricing} is $\Omega(\eps^{-3})$.
\end{theorem}

\begin{remark}
This result is most relevant to the work \cite{LSTW23soda}, which shows that:\\
(i)~For \textit{any number of general distributions}, the query complexity of {\UniformPricing} is $\tTheta(\eps^{-3})$.\\
(ii)~For \textit{a single regular distribution}, the query complexity of {\UniformPricing} is $\tTheta(\eps^{-2})$.\\
Consequently, \Cref{thm:two-regular} complements \cite{LSTW23soda} by showing a more thorough picture: ``a single regular distribution'' is a quite singular case --- even the \textit{minimal generalization} to ``two regular distributions'' will increase the query complexity from $\tTheta(\eps^{-2})$ to the general-case bound $\tTheta(\eps^{-3})$.
\end{remark}

In the remainder of this section, we will establish \Cref{thm:two-regular}. Without loss of generality, we consider a sufficiently small $\eps \in (0, \frac{1}{16})$ and a sufficiently large $K \defeq \lfloor \frac{1}{2}\eps^{-1} \rfloor \ge 8$ throughout. The entire proof takes two steps. Firstly, we present in \Cref{subsec:regular:construction} our lower-bound construction, including one base instance $\bF^{*}$ and $K$ hard instances $\{\bF^{i}\}_{i \in [K]}$. Afterward, we present in \Cref{subsec:regular:analysis} our lower-bound analysis, including
(i)~a reduction from the original \textit{pricing} problem to a new \textit{instance-identification} problem and
(ii)~a proof of a matching lower bound $\Omega(\eps^{-3})$ for the new problem. (Notably, later in \Cref{sec:three-MHR}, we will extend these lower-bound construction and analysis to the setting with MHR distributions.)

\subsection{Lower Bound Construction}
\label{subsec:regular:construction}

Consider two parameters $t \in [\frac{1}{2} + \eps, 1]$ and $s = s(t) \defeq t - \eps$. We define two parametric CDF's $F_{1}^{*t}$ and $F_{2}^{*t}$; see \Cref{fig:two-regular:1} for a diagram:
\begin{align*}
    F_{1}^{*t}(x)
    & ~\defeq~
    \begin{cases}
        \frac{x}{x + \frac{t}{3t - 1}}, & x \in [0, t]\\
        \frac{x - \frac{1}{3}}{x},  & x \in (t, 1]
    \end{cases},\\
    F_{2}^{*t}(x)
    & ~\defeq~
    \begin{cases}
        0, & x \in [0, \frac{1}{3}]\\
        \frac{x - \frac{1}{3}}{x} \cdot \frac{x + \frac{t}{3t - 1}}{x},  & x \in (\frac{1}{3}, t]\\
        1, & x \in (t, 1]
    \end{cases}.
\end{align*}
(We will verify the regularity of $F_{1}^{*t}$ and $F_{2}^{*t}$ later in \Cref{lem:two-regular:regular}.)
In regard to {\UniformPricing}, the first-order CDF $F^{*t}(x) \defeq F_{1}^{*t}(x) \cdot F_{2}^{*t}(x)$ and the revenue function $R^{*t}(x) \defeq x \cdot (1 - F^{*t}(x))$ are given as follows; see \Cref{fig:two-regular:3} for a diagram:
\begin{align*}
    F^{*t}(x)
    & ~=~
    \begin{cases} 
        0, & x \in [0, \frac{1}{3}]\\
        \frac{x - \frac{1}{3}}{x}, & x \in (\frac{1}{3}, 1]
    \end{cases},\\
    R^{*t}(x)
    & ~=~
    \begin{cases} 
        x, & x \in [0, \frac{1}{3}]\\
        \frac{1}{3}, & x \in (\frac{1}{3}, 1]
    \end{cases}.
\end{align*}
These formulae $F^{*t}$ and $R^{*t}$ are independent of $t \in [\frac{1}{2} + \eps, 1]$; accordingly, we can define our base instance $\bF^{*} \defeq F_{1}^{*t} \otimes F_{2}^{*t}$ using any specific $t \in [\frac{1}{2} + \eps, 1]$.

To establish the desired lower bound $\Omega(\eps^{-3})$, we modify the second parametric CDF $F_{2}^{*t}$ into another parametric CDF $F_{2}^{t}$ on the interval $(s, t] = (t - \eps, t]$, as follows; see \Cref{fig:two-regular:2} for a diagram. (In contrast, the first parametric CDF $F_{1}^{*t}$ keeps the same.)
\begin{align*}
    F_{2}^{t}(x) ~\defeq~
    \begin{cases}
        1 - \frac{(1 - F_{2}^{*t}(s))^{2}}{(x - s) \cdot f_{2}^{*t}(s) + (1 - F_{2}^{*t}(s))}
        ~=~ 1 - \frac{(1 - F_{2}^{*t}(s))^{2} / f_{2}^{*t}(s)}{x - \phi_{2}^{*t}(s)}, & x \in (s, t]\\
        F_{2}^{*t}(x), & x \notin (s, t]
    \end{cases}.
\end{align*}
(I.e., only pricing queries within the modification interval $(s, t]$ can help identify $F_{2}^{t}$.)
Then, the modified first-order CDF $F^{t}(x) \defeq F_{1}^{*t}(x) \cdot F_{2}^{t}(x)$ and the modified revenue function $R^{t}(x) \defeq x \cdot (1 - F^{t}(x))$ follow accordingly; see \Cref{fig:two-regular:4} for a diagram.

\begin{remark}
The modified parametric CDF $F_{2}^{t}$ is defined such that \textit{``over the interval $x \in (s, t]$, the corresponding virtual value function $\phi_{2}^{t}(x)$ is constant $= \phi_{2}^{*t}(s)$''}; below, the proof of \Cref{lem:two-regular:regular} will show this explicitly.
This induces an ordinary differential equation (ODE), and solving it under the boundary condition $\lim_{x \to s^{+}} F_{2}^{t}(x) = F_{2}^{*t}(s)$ (so as to preserve the continuity at $x = s$) gives the above defining formula.
\end{remark}

\def\ct{0.7}
\def\ce{0.1}
\def\cs{0.6}
\def\fs{(2*\ct-\cs)/(3*\cs^3*(3*\ct-1))}
\def\Fs{(1-1/(3*\cs))*(1+\ct/((3*\ct-1)*\cs))}
\begin{figure}[t]
	\centering
	\begin{minipage}{0.45\linewidth}
	\centering
	\subfigure[``Base'' (Second) CDF $F_{2}^{*t}$\label{fig:two-regular:1}]{
	\begin{tikzpicture}[scale=1]
		\begin{axis}[axis lines=middle,
			xlabel=$\scriptstyle x$,
			ylabel=$\scriptstyle F_2^{*t}$,
			xmin=0, xmax=1,
			ymin=0, ymax=1.1,
			tick label style={font=\small},
			scale only axis,
            width=0.84\textwidth,
            height=0.6\textwidth,
			legend style={font=\small,legend pos= north east,}]
			\addplot+[no marks,blue,domain=0:1/3,samples=20, thick] {0};
			\addplot+[no marks,blue,domain=1/3:\ct,samples=20, thick] {(1-1/(3*x))*(1+\ct/((3*\ct-1)*x))};
			\addplot+[no marks,blue,domain=\ct:1,samples=20, thick] {1};
		\end{axis}
	\end{tikzpicture}}
	\end{minipage}
	\hfill
	\begin{minipage}{0.45\linewidth}
	\centering
	\subfigure[``Hard'' (Second) CDF $F_2^t$\label{fig:two-regular:2}]{
	\begin{tikzpicture}[scale=1]
		\begin{axis}[axis lines=middle,
			xlabel=$\scriptstyle x$,
			ylabel=$\scriptstyle F_2^{t}$,
			xmin=0, xmax=1,
			ymin=0, ymax=1.1,
			tick label style={font=\small},
			scale only axis,
            width=0.84\textwidth,
            height=0.6\textwidth,
			legend style={font=\small,legend pos= north east,}]
			\addplot+[no marks,blue,domain=0:1/3,samples=20, thick] {0};
			\addplot+[no marks,blue,domain=1/3:\cs,samples=20, thick] {(1-1/(3*x))*(1+\ct/((3*\ct-1)*x))};
			\addplot+[no marks,dotted,domain=\cs:\ct,samples=20] {(1-1/(3*x))*(1+\ct/((3*\ct-1)*x))};
			\addplot+[no marks,red,domain=\cs:\ct,samples=20, thick] {1-(1-\Fs)^2/((x-\cs)*\fs+1-\Fs)};
			\addplot+[no marks,blue,domain=\ct:1,samples=20, thick] {1};
		\end{axis}
    \end{tikzpicture}}
    \end{minipage}
    
    \vspace{.4cm}
    \begin{minipage}{0.45\linewidth}
	\centering
	\subfigure[``Base'' Revenue Function $R^{*t}$\label{fig:two-regular:3}]{
		\begin{tikzpicture}[scale=1]
			\begin{axis}[axis lines=middle,
				xlabel=$\scriptstyle x$,
				ylabel=$\scriptstyle R^{*t}$,
				xmin=0, xmax=1,
				ymin=0, ymax=0.35,
				tick label style={font=\small},
				scale only axis,
                width=0.84\textwidth,
                height=0.6\textwidth,
				legend style={font=\small,legend pos= north east,}]
				\addplot+[no marks,blue,domain=0:1/3,samples=20, thick] {x};
				\addplot+[no marks,blue,domain=1/3:1,samples=20, thick] {1/3};
			\end{axis}
		\end{tikzpicture}
	}
	\end{minipage}
	\hfill
	\begin{minipage}{0.45\linewidth}
	\centering
	\subfigure[``Hard'' Revenue Function $R^{t}$\label{fig:two-regular:4}]{
	\begin{tikzpicture}[scale=1]
		\begin{axis}[axis lines=middle,
			xlabel=$\scriptstyle x$,
			ylabel=$\scriptstyle R^t$,
			xmin=0, xmax=1,
			ymin=0, ymax=0.35,
			tick label style={font=\small},
			scale only axis,
            width=0.84\textwidth,
            height=0.6\textwidth,
			legend style={font=\small,legend pos= north east,}]
			\addplot[no marks,blue,domain=0:1/3,samples=20, thick] {x};
			\addplot[no marks,blue,domain=1/3:\cs,samples=20, thick] {1/3};
			\addplot[no marks,dotted,domain=\cs:\ct,samples=20, thick] {1/3};
			\addplot[no marks,red,domain=\cs:\ct,samples=20, thick] {x*(1-(x/(x+\ct/(3*\ct-1)))*(1-(1-\Fs)^2/((x-\cs)*\fs+1-\Fs)) )};
			\addplot[no marks,blue,domain=\ct:1,samples=20, thick] {1/3};
		\end{axis}
	\end{tikzpicture}}
	\end{minipage}
	\caption{Diagrams for the lower-bound construction\ignore{ with $t=0.7$, $\eps=0.1$} in the ``two regular distributions'' setting.\label{fig:two-regular}}
\end{figure}

The following \Cref{lem:two-regular:regular} verifies the regularity of all considered CDF's.

\begin{lemma}[Regularity]
\label{lem:two-regular:regular}
Given any $t \in [\frac{1}{2} + \eps, 1]$, all of $F_{1}^{*t}$, $F_{2}^{*t}$, and $F_{2}^{t}$ are well-defined regular CDF's.
\end{lemma}

\begin{proof}
The first function $F_{1}^{*t}$ is \textit{continuous} at $x = t$ (since $\lim_{x \to t^{+}} F_{1}^{*t}(x) = \frac{t - \frac{1}{3}}{t} = F_{1}^{*t}(t)$), has a \textit{nonnegative} derivative function $f_{1}^{*t}$ (as follows), takes values between $0 = F_{1}^{*t}(0) \le F_{1}^{*t}(1) = \frac{2}{3}$, and has a \textit{nondecreasing} virtual value function $\phi_{1}^{*t}$ (as follows).
Given these, $F_{1}^{*t}$ is a well-defined regular CDF.
\begin{align*}
    f_{1}^{*t}(x)
    &\textstyle
    ~=~ F_{1}^{*t}{}'(x) ~=~
    \begin{cases}
        \frac{\frac{t}{3t - 1}}{(x + \frac{t}{3t - 1})^{2}}, & x \in [0, t]\\
        \frac{1}{3x^{2}},  & x \in (t, 1]
    \end{cases},\\
    \phi_{1}^{*t}(x)
    &\textstyle
    ~=~ x - \frac{1 - F_{1}^{*t}(x)}{f_{1}^{*t}(x)} ~=~
    \begin{cases}
        -\frac{t}{3t - 1}, & x \in [0, t]\\
        0,  & x \in (t, 1]
    \end{cases}.
\end{align*}

The second function $F_{2}^{*t}$ is $(\frac{1}{3}, t]$-supported, is \textit{continuous} at the two endpoints $\lim_{x \to (\frac{1}{3})^{+}} F_{2}^{*t}(x) = 0 = F_{2}^{*t}(\frac{1}{3})$ and $\lim_{x \to t^{+}} F_{2}^{*t}(x) = 1 = F_{2}^{*t}(t)$, has a \textit{nonnegative} derivative function $f_{2}^{*t}$ (as follows), takes values between $0 = F_{2}^{*t}(\frac{1}{3}) \le F_{2}^{*t}(t) = 1$, and has a \textit{nondecreasing} virtual value function $\phi_{2}^{*t}$ (as follows).
Given these, $F_{2}^{*t}$ is a well-defined regular CDF.
\begin{align*}
    f_{2}^{*t}(x)
    &\textstyle ~=~ F_{2}^{*t}{}'(x)
    ~=~ \frac{2t - x}{3x^{3} \cdot (3t - 1)},
    && x \in (\tfrac{1}{3}, t],\\
    \phi_{2}^{*t}(x)
    &\textstyle ~=~ x - \frac{1 - F_{2}^{*t}(x)}{f_{2}^{*t}(x)}
    ~=~ \frac{t \cdot x}{2t - x},
    && x \in (\tfrac{1}{3}, t].
\end{align*}

The modified function $F_{2}^{t}$ is $(\frac{1}{3}, t]$-supported --- with different defining formulae on the sub-intervals $(\frac{1}{3}, s]$ and $(s, t]$ --- is \textit{continuous} at the division point $\lim_{x \to s^{+}} F_{2}^{t}(x) = \ignore{\lim_{x \to s^{+}} \Big(1 - \frac{(1 - F_{2}^{*t}(s))^{2}}{(x - s) \cdot f_{2}^{*t}(s) + (1 - F_{2}^{*t}(s))}\Big) =} F_{2}^{*t}(s) = F_{2}^{t}(s)$, has a \textit{nonnegative} derivative function $f_{2}^{t}$ (as follows), takes values between $0 = F_{2}^{t}(\frac{1}{3}) \le F_{2}^{t}(t) \le 1$, and has a \textit{nondecreasing} virtual value function $\phi_{2}^{t}$ (as follows).
Given these, $F_{2}^{t}$ is a well-defined regular CDF.
\begin{align*}
    f_{2}^{t}(x)
    & ~=~ F_{2}^{t}{}'(x) ~=~
    \begin{cases}
        \frac{(1 - F_{2}^{*t}(s))^{2} / f_{2}^{*t}(s)}{(x - \phi_{2}^{*t}(s))^{2}},
        & x \in (s, t]\\
        f_{2}^{*t}(x),
        & x \notin (s, t]
    \end{cases}, \\
    \phi_{2}^{t}(x)
    & ~=~ x - \tfrac{1 - F_{2}^{t}(x)}{f_{2}^{t}(x)} ~=~
    \begin{cases}
        x - \frac{\frac{(1 - F_{2}^{*t}(s))^{2} / f_{2}^{*t}(s)}{x - \phi_{2}^{*t}(s)}}{\frac{(1 - F_{2}^{*t}(s))^{2} / f_{2}^{*t}(s)}{(x - \phi_{2}^{*t}(s))^{2}}}
        ~=~ \phi_{2}^{*t}(s),
        & x \in (s, t]\\
        \phi_{2}^{*t}(x),
        & x \notin (s, t]
    \end{cases}.
\end{align*}

This finishes the proof of \Cref{lem:two-regular:regular}.
\end{proof}

In addition, the following \Cref{lem:two-regular:first-order,lem:two-regular:revenue} will be useful for our lower-bound analysis in \Cref{subsec:regular:analysis}.

\begin{lemma}[First-Order CDF's]
\label{lem:two-regular:first-order}
$0 \le F^{*t}(x) - F^{t}(x) \le \frac{4}{3}\eps$ for $x \in (s, t]$, while $F^{*t}(x) = F^{t}(x)$ for $x \notin (s, t]$.
\end{lemma}

\begin{proof}
Following the proof of \Cref{lem:two-regular:regular}, we have $\phi_{2}^{*t}(x) \ge \phi_{2}^{*t}(s) = \phi_{2}^{t}(x) \iff -\frac{f_{2}^{*t}(x)}{1 - F_{2}^{*t}(x)} \le -\frac{f_{2}^{t}(x)}{1 - F_{2}^{t}(x)} \iff \frac{\d}{\d x}\ln(1 - F_{2}^{*t}(x)) \le \frac{\d}{\d x}\ln(1 - F_{2}^{t}(x))$, for $x \in (s, t]$.
Together with the boundary condition $F_{2}^{*t}(s) = F_{2}^{t}(s)$, we deduce that $F_{2}^{*t}(x) - F_{2}^{t}(x) \ge 0 \implies F^{*t}(x) - F^{t}(x) = F_{1}^{*t}(x) \cdot \big(F_{2}^{*t}(x) - F_{2}^{t}(x)\big) \ge 0$, for $x \in (s, t]$.
Also, we have $f^{*t}(x) - f^{t}(x) \le f^{*t}(x) = \frac{1}{3x^{2}} \le \frac{4}{3}$, for $x \in (s, t] \subseteq [\frac{1}{2}, 1]$, where the first step uses the modified first-order PDF $f^{t}$'s nonnegativity. Together with the boundary condition $F^{*t}(s) = F^{t}(s)$, we deduce that $F^{*t}(x) - F^{t}(x) \le \frac{4}{3} \cdot (x - s) \le \frac{4}{3} \cdot (t - s) = \frac{4}{3}\eps$, for $x \in (s, t]$.

The second part ``$F^{*t}(x) = F^{t}(x)$ for $x \notin (s, t]$'' is obvious. This finishes the proof of \Cref{lem:two-regular:first-order}.
\end{proof}

\begin{lemma}[Revenue Functions]
\label{lem:two-regular:revenue}
$R^{t}(t) \ge \frac{1}{3} + \frac{1}{18}\eps$, while $R^{t}(x) \le \frac{1}{3}$ for $x \notin (s, t]$.
\end{lemma}

\begin{proof}
Following the proof of \Cref{lem:two-regular:regular}, we have
$F_{1}^{*t}(t)
= \frac{t - \frac{1}{3}}{t}$,\quad
$F_{2}^{*t}(s)
= \frac{s - \frac{1}{3}}{s} \cdot \frac{s + \frac{t}{3t - 1}}{s}
= 1 - \frac{\eps}{3 \cdot (t - \eps)^{2} \cdot (3t - 1)}$,\quad and\quad
$f_{2}^{*t}(s)
= \frac{2t - s}{3s^{3} \cdot (3t - 1)}
= \frac{t + \eps}{3 \cdot (t - \eps)^{3} \cdot (3t - 1)}$.

Given that $R^{*t}(t) = \frac{1}{3}$, we can deduce the first part $R^{t}(t) \ge \frac{1}{3} + \frac{1}{18}\eps$, as follows:
\begin{align*}
    R^{t}(t) - R^{*t}(t)
    &\textstyle ~=~ t \cdot F_{1}^{*t}(t) \cdot (F_{2}^{*t}(t) - F_{2}^{t}(t))\\
    &\textstyle ~=~ t \cdot F_{1}^{*t}(t) \cdot \frac{(1 - F_{2}^{*t}(s))^{2}}{(t - s) \cdot f_{2}^{*t}(s) + (1 - F_{2}^{*t}(s))}\\
    &\textstyle ~=~ t \cdot \frac{3t - 1}{3t} \cdot \frac{\big(\frac{\eps}{3 \cdot (t - \eps)^{2} \cdot (3t - 1)}\big)^{2}}{\eps \cdot \frac{t + \eps}{3 \cdot (t - \eps)^{3} \cdot (3t - 1)} + \frac{\eps}{3 \cdot (t - \eps)^{2} \cdot (3t - 1)}}\\
    &\textstyle ~=~ \frac{\eps}{18t \cdot (t - \eps)}\\
    &\textstyle ~\ge~ \frac{1}{18}\eps.
\end{align*}
Here the first two steps substitutes the defining formulae of $R^{t}(t)$, $R^{*t}(t)$, and $F_{2}^{t}(t)$; note that $F_{2}^{*t}(t) = 1$.
The third step substitutes the above formulae of $F_{1}^{*t}(t)$, $F_{2}^{*t}(s)$, and $f_{2}^{*t}(s)$.
The fourth step rearranges the equation.
And the last step holds since $\eps \in (0, \frac{1}{16})$ and $t \in [\frac{1}{2} + \eps, 1]$.

The second part ``$R^{t}(x) = R^{*t}(x) \le \frac{1}{3}$ for $x \notin (s, t]$'' is obvious. This finishes the proof of \Cref{lem:two-regular:revenue}.
\end{proof}

To obtain the desired lower bound $\Omega(\eps^{-3})$, we choose a sequence of \textit{disjoint} modification intervals\footnote{Since $K = \lfloor \frac{1}{2}\eps^{-1} \rfloor$, all parameters $\{t^{i}\}_{i \in [K]}$ belong to the interval $[\frac{1}{2} + \eps, 1]$, satisfying the premises of \Cref{lem:two-regular:regular,lem:two-regular:revenue}.} $(s^{i}, t^{i}] \defeq (\frac{1}{2} + i \eps - \eps, \frac{1}{2} + i \eps]$ and, thus, obtain our hard instances $\bF^{i} \defeq F_{1}^{*t^{i}} \otimes F_{2}^{t^{i}}$ for $i \in [K]$.

\subsection{Lower Bound Analysis}
\label{subsec:regular:analysis}

Consider a specific pricing algorithm $\+A$ that performs well on all hard instances $\{\bF^{i}\}_{i \in [K]}$: in any possibility $i \in [K]$, it always outputs a $\frac{1}{20}\eps$-approximately optimal price $p^{\+A}$ with probability $\ge \frac{2}{3}$:
\begin{align*}
    \textstyle
    R^{i}(p^{\+A})
    ~\ge~ \max_{p \in [0, 1]} R^{i}(p) - \frac{1}{20}\eps.
\end{align*}
(In contrast, no performance guarantee for the base instance $\bF^{*}$ is needed.)
Based on this, we can develop a ``pricing-to-identification'' reduction for all base/hard instances $\{\bF^{*}\} \cup \{\bF^{i}\}_{i \in [K]}$, i.e., another identification algorithm $\+B^{\+A}$ with \textit{exactly the same number of pricing queries}:
\begin{itemize}
    \item Run $\+A$ on an unknown instance $\bF$ (promised to be one of $\{\bF^{*}\} \cup \{\bF^{i}\}_{i \in [K]}$), getting a price $p^{\+A}$.
    
    \item If $p^{\+A} \in (s^{i}, t^{i}]$ for some $i \in [K]$, output $\bF^{i}$;
    
    \item Otherwise, output $\bF^{*}$.
\end{itemize}
Clearly, $\+B^{\+A}$ can identify all hard instances $\{\bF^{i}\}_{i \in [K]}$:\footnote{I.e., this directly follows from a combination of \Cref{lem:two-regular:revenue}, that the modification intervals $(s^{i}, t^{i}]$ are disjoint, and $\+A$'s performance guarantees.}
in any possibility $i \in [K]$, it always successfully outputs $\bF^{i}$ with probability $\ge \frac{2}{3}$.
(Again, no performance guarantee for the base instance $\bF^{*}$ is needed.)

In the rest of this section, we consider a specific identification algorithm $\+B$; denote by $T$ the number of pricing queries it makes and, in particular, $T^{i}$ for $i \in [K]$ the number of pricing queries it makes within the index-$i$ modification interval $(s^{i}, t^{i}]$. Also, denote by $\bb{P}^{*}[\cdot]$ or $\bb{P}^{i}[\cdot]$ the probabilities in each possibility $i \in [K]$; likewise for the expectations $\bb{E}^{*}[\cdot]$ or $\bb{E}^{i}[\cdot]$.

The following \Cref{lem:two-regular:identify} lower-bounds the query complexity of an identification algorithm $\+B$.

\begin{lemma}[Identification Lower Bounds]
\label{lem:two-regular:identify}
To identify hard instances $\{\bF^{i}\}_{i \in [K]}$ each with probability $\ge \frac{2}{3}$, an identification algorithm $\+B$ makes at least $\bb{E}^{*}[T] = \Omega(\eps^{-3})$ many pricing queries on the base instance $\bF^{*}$ (in expectation over the randomness of both $\+B$ itself and $\bF^{*}$).
\end{lemma}

\begin{proof}
Consider the base instance $\bF^{*}$ and a specific hard instance $\bF^{i}$.
As mentioned, only pricing queries within the corresponding modification interval $(s^{i}, t^{i}]$ can help identify $\bF^{i}$.

Recall that $\DKL(p, q) = p \ln(\frac{p}{q}) + (1 - p) \ln(\frac{1 - p}{1 - q})$ denotes the KL divergence between two Bernoulli distributions with parameters $p, q \in [0, 1]$.
For $x \in (s^{i}, t^{i}] \subseteq [\frac{1}{2}, 1]$, we have $0 \le F^{*}(x) - F^{i}(x) \le \frac{4}{3}\eps \le \frac{1}{12}$ (\Cref{lem:two-regular:first-order}) and $F^{*}(x) = \frac{3x - 1}{3x} \in [\frac{1}{3}, \frac{2}{3}]$, so \Cref{cla:DKL-UB} is applicable and gives
\begin{align*}
    \textstyle
    \DKL\big(F^{*}(x),\ F^{i}(x)\big)
    ~\le~ 3 \cdot (\frac{4}{3}\eps)^{2}
    ~=~ \frac{16}{3}\eps^{2}.
\end{align*}
Then, regarding the event $\@E^{i} \defeq \{\text{$\+B$ outputs $\bF^{i}$}\}$, we know from \Cref{lem:kl-bound} that
\begin{align*}
    \textstyle
    \DKL\big(\bb{P}^{*}[\@E^{i}],\ \bb{P}^{i}[\@E^{i}]\big)
    ~\le~ \DKL\big(F^{*}(x),\ F^{i}(x)\big) \cdot \bb{E}^{*}[T^{i}]
    ~\le~ \frac{16}{3}\eps^{2} \cdot \bb{E}^{*}[T^{i}].
\end{align*}
By enumerating all $i \in [K]$, we can upper-bound the sum $\sum_{i \in [K]} \DKL(\bb{P}^{*}[\@E^{i}],\ \bb{P}^{i}[\@E^{i}])$ as follows:
\begin{align*}
    \textstyle
    \sum_{i \in [K]} \DKL\big(\bb{P}^{*}[\@E^{i}],\ \bb{P}^{i}[\@E^{i}]\big)
    ~\le~ \sum_{i \in [K]} \frac{16}{3}\eps^{2} \cdot \bb{E}^{*}[T^{i}]
    ~\le~ \frac{16}{3}\eps^{2} \cdot \bb{E}^{*}[T].
\end{align*}
Here the last step uses the linearity of expectations and that $\sum_{i \in [K]} T^{i} \le T$ (in any possible outcome).

Moreover, since the KL divergence $\DKL(p, q)$ is a convex function (\Cref{cla:DKL-convex}), using Jensen's inequality (\Cref{cla:Jensen}), we can lower-bound the sum $\sum_{i \in [K]} \DKL(\bb{P}^{*}[\@E^{i}],\ \bb{P}^{i}[\@E^{i}])$ as follows:
\begin{align*}
    \textstyle
    \sum_{i \in [K]} \DKL\big(\bb{P}^{*}[\@E^{i}],\ \bb{P}^{i}[\@E^{i}]\big)
    &\textstyle ~\ge~ K \cdot \DKL\Big(\frac{\sum_{i \in [K]} \bb{P}^{*}[\@E^{i}]}{K},\ \frac{\sum_{i \in [K]} \bb{P}^{i}[\@E^{i}]}{K}\Big)\\
    &\textstyle ~\ge~ K \cdot \DKL(\frac{1}{8}, \frac{2}{3})\\
    &\textstyle ~\ge~ \frac{1}{2}K.
\end{align*}
Here the second step uses ``$\{\@E^{i}\}_{i \in [K]}$ are disjoint'' $\implies \frac{\sum_{i \in [K]} \bb{P}^{*}[\@E^{i}]}{K} \le \frac{1}{K}\le \frac{1}{8}$ and the premise of the lemma ``$\bb{P}^{i}[\@E^{i}] \ge \frac{2}{3}$ for $i \in [K]$''.
And the last step uses $\DKL(\frac{1}{8}, \frac{2}{3}) \approx 0.6352$.

Combining the above two equations directly gives $\bb{E}^{*}[T] \ge \frac{3}{32}\eps^{-2} \cdot K \ge \frac{21}{512}\eps^{-3}$, where the last step uses $\eps \in (0, \frac{1}{16}) \implies K = \lfloor \frac{1}{2}\eps^{-1} \rfloor \ge \frac{7}{16}\eps^{-1}$. This finishes the proof of \Cref{lem:two-regular:identify}.
\end{proof}

Finally, we can translate the query complexity lower bound of an identification algorithm (\Cref{lem:two-regular:identify}) into that of a pricing algorithm (\Cref{thm:two-regular}).

\begin{proof}[Proof of \Cref{thm:two-regular}]
Incorporate the ``pricing-to-identification'' reduction into \Cref{lem:two-regular:identify}: if a pricing algorithm $\+A$ always outputs a $\frac{1}{20}\eps$-approximately optimal price $p^{\+A}$ with probability $\ge \frac{2}{3}$, then it makes at least $\Omega(\eps^{-3})$ many pricing queries on the base instance $\bF^{*}$.

Scaling the parameter $\eps \in (0, \frac{1}{16})$ by a factor of $\frac{1}{20}$ finishes the proof of \Cref{thm:two-regular}.
\end{proof}

\section{\texorpdfstring{$\Omega(\eps^{-3})$ Lower Bound for Three MHR Distributions}{}}
\label{sec:three-MHR}

In this section, we investigate the query complexity of {\UniformPricing} in the setting with \textit{MHR} distributions. Specifically, we will establish the following \Cref{thm:three-MHR}.

\begin{theorem}
\label{thm:three-MHR}
For three (or more) MHR distributions, the query complexity of {\UniformPricing} is $\Omega(\eps^{-3})$.
\end{theorem}

\begin{remark}
This result is most relevant to the work \cite{LSTW23soda} (as well as \Cref{thm:two-regular}), which shows that:\\
(i)~For \textit{any number of general distributions}, the query complexity of {\UniformPricing} is $\tTheta(\eps^{-3})$.\\
(ii)~For \textit{a single MHR distribution}, the query complexity of {\UniformPricing} is $\tTheta(\eps^{-2})$.\\
Consequently, \Cref{thm:three-MHR} complements \cite{LSTW23soda} (as well as \Cref{thm:two-regular}) by showing a more thorough picture: ``a single MHR distribution'' is a quite singular case --- even a \textit{minor generalization} to ``three MHR distributions'' will increase the query complexity from $\tTheta(\eps^{-2})$ to the general-case bound $\tTheta(\eps^{-3})$.

We are left with the intermediate case ``two MHR distributions''. Unfortunately, for this case we can only establish a query complexity lower bound of $\Omega(\eps^{-5 / 2})$, which is deferred to \Cref{sec:two-MHR}. It is interesting to close the gap between the best known bounds $\Omega(\eps^{-5 / 2})$ and $O(\eps^{-3})$ for this case.
\end{remark}

In the remainder of this section, we will establish \Cref{thm:three-MHR}. Again, we consider a sufficiently small $\eps \in (0, \frac{1}{48})$ and a sufficiently large $K \defeq \lfloor \frac{1}{8}\eps^{-1} \rfloor \ge 6$ throughout. The proof naturally adapts the techniques in \Cref{sec:regular} (from the ``two regular distributions'' case) to the ``three MHR distributions'' case. Firstly, we present in \Cref{subsec:three-MHR:construction} the counterpart lower-bound construction, including one base instance $\bF^{*}$ and $K$ hard instances $\{\bF^{i}\}_{i \in [K]}$. Afterward, we present in \Cref{subsec:three-MHR:analysis} the counterpart lower-bound analysis.



Without ambiguity, we often reload the notations introduced in \Cref{sec:regular}.

\subsection{Lower Bound Construction}
\label{subsec:three-MHR:construction}

Consider two parameters $t \in [\frac{7}{8} + \eps, 1]$ and $s = s(t) \defeq t - \eps$. We define three parametric CDF's $F_{1}^{*t}$, $F_{2}^{*t}$, and $F_{3}^{*t}$:
\begin{align*}
    F_{1}^{*t}(x)
    &\textstyle ~\defeq~ 1 - (\frac{3}{4})^{x},
    &&\hspace{-4.87cm}\textstyle x \in [0, 1],\\
    F_{2}^{*t}(x)
    & ~\defeq~
    \begin{cases}
        \frac{1 - \frac{3}{4t}}{1 - (\frac{3}{4})^{t}} \cdot \frac{x}{t},
        &\hspace{1.3cm} x \in [0, t]\\
        \frac{1 - \frac{3}{4x}}{1 - (\frac{3}{4})^{x}},
        &\hspace{1.3cm} x \in (t, 1]
    \end{cases},\\
    F_{3}^{*t}(x)
    & ~\defeq~
    \begin{cases}
        0 & x \in [0, \frac{3}{4}]\\
        \frac{1 - \frac{3}{4x}}{1 - (\frac{3}{4})^{x}} \cdot \frac{1 - (\frac{3}{4})^{t}}{1 - \frac{3}{4t}} \cdot \frac{t}{x}, & x \in (\frac{3}{4}, t]\\
        1, & x \in (t, 1]
    \end{cases}.
\end{align*}
(We will verify the MHR condition for these distributions later in \Cref{lem:three-MHR:MHR}.)
In regard to {\UniformPricing}, the first-order CDF $F^{*t}(x) \defeq F_{1}^{*t}(x) \cdot F_{2}^{*t}(x) \cdot F_{3}^{*t}(x)$ and the revenue function $R^{*t}(x) \defeq x \cdot (1 - F^{*t}(x))$ are given as follows:
\begin{align*}
    F^{*t}(x)
    & ~=~
    \begin{cases} 
        0, & x \in [0, \frac{3}{4}]\\
        \frac{x - \frac{3}{4}}{x}, & x \in (\frac{3}{4}, 1]
    \end{cases},\\
    R^{*t}(x)
    & ~=~
    \begin{cases} 
        x, & x \in [0, \frac{3}{4}]\\
        \frac{3}{4}, & x \in (\frac{3}{4}, 1]
    \end{cases}.
\end{align*}
These formulae $F^{*t}$ and $R^{*t}$ are independent of $t \in [\frac{7}{8} + \eps, 1]$; accordingly, we can define our base instance $\bF^{*} \defeq F_{1}^{*t} \otimes F_{2}^{*t} \otimes F_{3}^{*t}$ using any specific $t \in [\frac{7}{8} + \eps, 1]$.

To establish the desired lower bound $\Omega(\eps^{-3})$, we modify the third parametric CDF $F_{3}^{*t}$ into another parametric CDF $F_{3}^{t}$ on the interval $(s, t] = (t - \eps, t]$, as follows. (In contrast, the first and the second parametric CDF's $F_{1}^{*t}$ and $F_{2}^{*t}$ keep the same.)
\begin{align*}
    F_{3}^{t}(x) ~\defeq~
    \begin{cases}
        1 - (1 - F_{3}^{*t}(s)) \cdot e^{-\frac{f_{3}^{*t}(s)}{1 - F_{3}^{*t}(s)} \cdot (x - s)}
        ~=~ 1 - (1 - F_{3}^{*t}(s)) \cdot e^{-h_{3}^{*t}(s) \cdot (x - s)}, & x \in (s, t]\\
        F_{3}^{*t}(x), & x \notin (s, t]
    \end{cases}.
\end{align*}
(I.e., only pricing queries within the modification interval $(s, t]$ can help identify $F_{3}^{t}$.)
Then, the modified first-order CDF $F^{t}(x) \defeq F_{1}^{*t}(x) \cdot F_{2}^{*t}(x) \cdot F_{3}^{t}(x)$ and the modified revenue function $R^{t}(x) \defeq x \cdot (1 - F^{t}(x))$ follow accordingly.

\begin{remark}
The modified parametric CDF $F_{3}^{t}$ is defined such that \textit{``over the interval $x \in (s, t]$, the corresponding hazard rate function $h_{3}^{t}(x)$ is constant $= h_{3}^{*t}(s)$''}; below, the proof of \Cref{lem:three-MHR:MHR} will show this explicitly.
This induces an ODE, and solving it under the boundary condition $\lim_{x \to s^{+}} F_{3}^{t}(x) = F_{3}^{*t}(s)$ (so as to preserve the continuity at $x = s$) gives the above defining formula.
\end{remark}

The following \Cref{lem:three-MHR:MHR} verifies the MHR condition for all considered CDF's.

\begin{lemma}[MHR]
\label{lem:three-MHR:MHR}
Given any $t \in [\frac{7}{8} + \eps, 1]$, all of $F_{1}^{*t}$, $F_{2}^{*t}$, $F_{3}^{*t}$, and $F_{3}^{t}$ are well-defined MHR CDF's.
\end{lemma}

\begin{proof}
Let $a \defeq \ln(\frac{4}{3}) \approx 0.2877$ and $b = b(t) \defeq \frac{1 - (3 / 4)^{t}}{t - 3 / 4} \cdot t^{2}$ for notational brevity.

The first function $F_{1}^{*t}(x) = 1 - (\frac{3}{4})^{x}$ for $x \in [0, 1]$ is clearly a well-defined MHR CDF.

The second function $F_{2}^{*t}$ is \textit{continuous} at $x = t$ (since $\lim_{x \to t^{+}} F_{2}^{*t}(x) = \frac{t}{b} = F_{2}^{*t}(t)$), has a \textit{positive} derivative function $f_{2}^{*t}$ (\Cref{claim:three-MHR:MHR:f2}), takes values between $0 = F_{2}^{*t}(0) \le F_{2}^{*t}(1) = 1$, and has a \textit{increasing} hazard rate function $h_{2}^{*t}$ (\Cref{claim:three-MHR:MHR:h2}).
Given these, $F_{2}^{*t}$ is a well-defined MHR CDF.
\begin{align*}
    f_{2}^{*t}(x)
    &\textstyle
    ~=~ F_{2}^{*t}{}'(x) ~=~
    \begin{cases}
        \frac{1}{b}, & x \in [0, t]\\
        \frac{(\frac{4}{3})^{x} - 1 - a x \cdot (\frac{4}{3}x - 1)}{(1 - (\frac{3}{4})^{x})^{2}} \cdot \frac{3}{4x^{2}} \cdot (\frac{3}{4})^{x},  & x \in (t, 1]
    \end{cases},\\
    h_{2}^{*t}(x)
    &\textstyle
    ~=~ \frac{f_{2}^{*t}(x)}{1 - F_{2}^{*t}(x)} ~=~
    \begin{cases}
        \frac{1}{b - x}, & x \in [0, t]\\
        \frac{(\frac{4}{3})^{x} - 1 - a x \cdot (\frac{4}{3}x - 1)}{(1 - (\frac{3}{4})^{x}) \cdot (\frac{3}{4x} - (\frac{3}{4})^{x})} \cdot \frac{3}{4x^{2}} \cdot (\frac{3}{4})^{x},  & x \in (t, 1]
    \end{cases}.
\end{align*}

\begin{claim}
\label{claim:three-MHR:MHR:f2}
$f_{2}^{*t}(x) \ge 0$ for $x \in [0, 1]$.
\Comment{Later, \Cref{claim:three-MHR:MHR:f2} and its proof will be useful for \Cref{lem:two-MHR:MHR}.}
\end{claim}

\begin{proof}
For $x \in (t, 1]$, note that $f_{2}^{*t}(x) \ge 0 \iff y(x) \defeq (\frac{4}{3})^{x} - 1 - a x \cdot (\frac{4}{3}x - 1) \ge 0$. Even on the wider interval $[\frac{3}{4}, 1] \supseteq (t, 1]$, this function is \textit{concave}  $y''(x) = (\frac{4}{3})^{x} \cdot a^{2} - \frac{8}{3}a \le \frac{4}{3}a^{2} - \frac{8}{3}a \approx -0.6568 < 0$ and have \textit{positive} endpoints $y(\frac{3}{4}) = (\frac{4}{3})^{3 / 4} - 1 \approx 0.2408$ and $y(1) = \frac{1 - a}{3} \approx 0.2374$, which gives $y(x) \ge 0$ for $x \in [\frac{3}{4}, 1]$.

For $x \in [0, t]$, we trivially have $f_{2}^{*t}(x) = \frac{1}{b} \ge 0$.
This finishes the proof of \Cref{claim:three-MHR:MHR:f2}.
\end{proof}

\begin{claim}
\label{claim:three-MHR:MHR:h2}
$h_{2}^{*t}{}'(x) \ge 0$ for $x \in [0, 1]$.
\Comment{Later, \Cref{claim:three-MHR:MHR:h2} and its proof will be useful for \Cref{lem:two-MHR:MHR}.}
\end{claim}

\begin{proof}
For $x \in (t, 1]$, we have $h_{2}^{*t}(x) = \frac{N(x)}{D(x)}$, where
\begin{align*}
    N(x) &\textstyle ~\defeq~ \frac{3}{4x^{2}} - (a - \frac{3a}{4x} + \frac{3}{4x^{2}}) \cdot (\frac{3}{4})^{x},\\
    D(x) &\textstyle ~\defeq~ \frac{3}{4x} - (1 + \frac{3}{4x}) \cdot (\frac{3}{4})^{x} + (\frac{3}{4})^{2x}.
\end{align*}
Note that $h_{2}^{*t}{}'(x) \ge 0 \iff y(x) \defeq N'(x) \cdot D(x) - N(x) \cdot D'(x) \ge 0$; we would even prove this on the wider interval $x \in [\frac{3}{4}, 1]$.
By elementary algebra, we deduce that
\begin{align*}
    N'(x) &\textstyle ~=~ -\frac{3}{2x^{3}} + (a^{2} - \frac{3a^{2}}{4x} + \frac{3}{2x^{3}}) \cdot (\frac{3}{4})^{x},\\
    D'(x) &\textstyle ~=~ -\frac{3}{4x^{2}} + (a + \frac{3a}{4x} + \frac{3}{4x^{2}}) \cdot (\frac{3}{4})^{x} - 2a \cdot (\frac{3}{4})^{2x},\\
    y(x) &\textstyle ~=~ (\frac{3}{4})^{x} \cdot \big(a^{2} \cdot (1 - \frac{3}{4x}) \cdot (\frac{3}{4x} - (\frac{3}{4})^{2x}) + \frac{9}{16x^{4}} \cdot (1 - (\frac{3}{4})^{x})^{2} \cdot (\frac{8x \cdot (1 - a x)}{3} - (\frac{4}{3})^{x})\big).
\end{align*}
For $x \in [\frac{3}{4}, 1]$, it is easy to check $\big(1 \ge \frac{3}{4x}\big) \land \big(\frac{3}{4x} \ge (\frac{3}{4})^{2x}\big) \land \big(\frac{8x \cdot (1 - a x)}{3} \ge \frac{5x}{3} \ge (\frac{4}{3})^{x}\big)$, which gives $y(x) \ge 0$.

For $x \in [0, t]$, we trivially have $h_{2}^{*t}{}'(x) = \frac{1}{(b - x)^{2}} \ge 0$.

At the division point $x = t \in [\frac{7}{8} + \eps, 1]$, we can deduce $\lim_{x \to t^{+}} h_{2}^{*t}(x) \ge h_{2}^{*t}(t)$ as follows:
\begin{align*}
    \textstyle
    \lim_{x \to t^{+}} h_{2}^{*t}(x)
    ~\ge~ h_{2}^{*t}(t)
    &\textstyle \quad\iff\quad \frac{\frac{3}{4t^{2}} \cdot (1 - (\frac{3}{4})^{t}) - (1 - \frac{3}{4t}) \cdot (\frac{3}{4})^{t} \cdot a}{(1 - (\frac{3}{4})^{t}) \cdot (\frac{3}{4t} - (\frac{3}{4})^{t})}
    ~\ge~ \frac{1}{b - t}
    ~\equiv~ \frac{1 - \frac{3}{4t}}{(\frac{3}{4t} - (\frac{3}{4})^{t}) \cdot t}\\
    &\textstyle \quad\iff\quad (\frac{2}{3} - a t) \cdot (t - \frac{3}{4})
    + \big((\frac{3}{2} - t) \cdot (\frac{4}{3})^{t} - 1 + \frac{t}{3}\big)
    ~\ge~ 0.
\end{align*}
Even on the wider interval $t \in [\frac{7}{8}, 1]$, it is easy to check $\big(\frac{2}{3} \ge a t\big) \land \big(t \ge \frac{3}{4}\big) \land \big((\frac{3}{2} - t) \cdot (\frac{4}{3})^{t} - 1 + \frac{t}{3} \ge 0\big)$, which gives $\lim_{x \to t^{+}} h_{2}^{*t}(x) \ge h_{2}^{*t}(t)$.
This finishes the proof of \Cref{claim:three-MHR:MHR:h2}.
\end{proof}

The third function $F_{3}^{*t}$ is $(\frac{3}{4}, t]$-supported, is \textit{continuous} at the two endpoints $\lim_{x \to (\frac{3}{4})^{+}} F_{3}^{*t}(x) = 0 = F_{3}^{*t}(\frac{3}{4})$ and $\lim_{x \to t^{+}} F_{3}^{*t}(x) = 1 = F_{3}^{*t}(t)$, has a \textit{positive} derivative function $f_{3}^{*t}$ (\Cref{claim:three-MHR:MHR:f3}), takes values between $0 = F_{3}^{*t}(\frac{3}{4}) \le F_{3}^{*t}(t) = 1$, and has a \textit{increasing} hazard rate function $h_{3}^{*t}$ (\Cref{claim:three-MHR:MHR:h3}).
Given these, $F_{3}^{*t}$ is a well-defined MHR CDF.
\begin{align*}
    f_{3}^{*t}(x)
    &\textstyle ~=~ F_{3}^{*t}{}'(x)
    ~=~ \frac{(6 - 4x) - (\frac{3}{4})^{x} \cdot (4a \cdot x^{2} - (3a + 4) \cdot x + 6)}{(1 - (\frac{3}{4})^{x})^{2}} \cdot \frac{b}{4x^{3}},
    && x \in (\tfrac{3}{4}, t],\\
    h_{3}^{*t}(x)
    &\textstyle ~=~ \frac{f_{3}^{*t}(x)}{1 - F_{3}^{*t}(x)}
    ~=~ \frac{(6 - 4x) - (\frac{3}{4})^{x} \cdot (4a \cdot x^{2} - (3a + 4) \cdot x + 6)}{(\frac{1}{b} \cdot (1 - (\frac{3}{4})^{x}) - \frac{4x - 3}{4x^{2}}) \cdot (1 - (\frac{3}{4})^{x})} \cdot \frac{1}{4x^{3}},
    && x \in (\tfrac{3}{4}, t].
\end{align*}

\begin{claim}
\label{claim:three-MHR:MHR:f3}
$f_{3}^{*t}(x) \ge 0$ for $x \in (\frac{3}{4}, t]$.
\end{claim}

\begin{proof}
By elementary algebra, $f_{3}^{*t}(x) \ge 0 \iff y(x) \ge 0$.
\begin{align*}
    \textstyle
    y(x) ~\defeq~ (6 - 4x) - (\frac{3}{4})^{x} \cdot \big(4a \cdot x^{2} - (3a + 4) \cdot x + 6\big).
\end{align*}
By elementary algebra, we have
\begin{align*}
    \textstyle
    y'(x) ~=~ -4 + (\frac{3}{4})^{x} \cdot \big(4a^{2} \cdot x^{2} - (3a^{2} + 12a) \cdot x + (9a + 4)\big).
\end{align*}
Consider the parabola $z(x) \defeq 4a^{2} \cdot x^{2} - (3a^{2} + 12a) \cdot x + (9a + 4)$; it opens upward (since $4a^{2} > 0 $) and has an axis of symmetry $ x = \frac{12 + 3a}{8a} \approx 5.5888 > 1$.
Thus, even on the wider interval $x \in [\frac{3}{4}, 1]$, it is \textit{decreasing} and \textit{positive} $z(x) \ge z(1) = a^{2} - 3a + 4 \approx 3.2197 > 0$.
Even on the wider interval $x \in [\frac{3}{4}, 1]$, it is easy to see that $4a^{2} \cdot x^{2} - (3a^{2} + 12a) \cdot x + (9a + 4) \le 4a^{2} \cdot 1^{2} - (3a^{2} + 12a) \cdot \frac{3}{4} + (9a + 4) \approx 4.1448 < \frac{9}{2} \implies y'(x) \le -4 + (\frac{3}{4})^{3 / 4} \cdot \frac{9}{2} \approx -0.3733$ and, consequently, that $y(x)$ is \textit{decreasing} and \textit{positive} $y(x) \ge y(1) = \frac{1}{2} - \frac{3}{4}a \approx 0.2842$.

This finishes the proof of \Cref{claim:three-MHR:MHR:f3}.
\end{proof}

\begin{claim}
\label{claim:three-MHR:MHR:h3}
$h_{3}^{*t}{}'(x) \ge 0$ for $x \in (\frac{3}{4}, t]$.
\end{claim}

\begin{proof}
For $x \in (\frac{3}{4}, t]$, we have $h_{3}^{*t}(x) = \frac{N(x)}{D(x)}$, where
\begin{align*}
    N(x) &\textstyle ~\defeq~ \big(\frac{3 - 2x}{2x^{3}} - \frac{6 - (4 + 3a) \cdot x + 4a x^{2}}{4x^{3}} \cdot (\frac{3}{4})^{x}\big) \cdot \frac{x}{t - x},\\
    D(x) &\textstyle ~\defeq~ \big(\frac{1}{b} \cdot (1 - (\frac{3}{4})^{x}) - \frac{4x - 3}{4x^{2}}\big) \cdot (1 - (\frac{3}{4})^{x}) \cdot \frac{x}{t - x}.
\end{align*}

Firstly, we assert that $N(x)$ is an \textit{increasing} and \textit{positive} function on $x \in (\frac{3}{4}, t]$. By elementary algebra,
\begin{align*}
    N'(x)
    &\textstyle ~=~ (4x - 3) \cdot \frac{(4 - 2x) \cdot (1 - (\frac{3}{4})^{x}) - a x \cdot (\frac{3}{4})^{x}}{4x^{3} \cdot (t - x)^{2}}
    ~+~ (4x - 3) \cdot \frac{a^{2} x^{2} \cdot (1 - x) \cdot (\frac{3}{4})^{x}}{4x^{3} \cdot (t - x)^{2}}\\
    &\textstyle \phantom{~=~} ~+~ (1 - t) \cdot \frac{(12 - 4x) \cdot (1 - (\frac{3}{4})^{x}) + a x \cdot (4x - 3) \cdot (1 - a x) \cdot (\frac{3}{4})^{x}}{4x^{3} \cdot (t - x)^{2}}.
\end{align*}
Here each of these three summands is \textit{positive}, given that $t \in [\frac{7}{8} + \eps, 1]$, $x \in (\frac{3}{4}, t]$, and $a = \ln(\frac{4}{3}) \approx 0.2877$; for the first summand specifically, even on the wider interval $x \in [0, 1]$, we deduce
$(\frac{3}{4})^{x} \le 1 - \frac{x}{4} \le 1 \implies (4 - 2x) \cdot (1 - (\frac{3}{4})^{x}) - a x \cdot (\frac{3}{4})^{x}
\ge (4 - 2x) \cdot \frac{x}{4} - a x
\ge (4 - 2x) \cdot \frac{x}{4} - \frac{x}{2}
= \frac{x \cdot (1 - x)}{2} \ge 0$.

The monotonicity of $N(x)$ together with $N(\frac{3}{4}) = \frac{4 / 3 - (4 / 3)^{1 / 4}}{t - 3 / 4} \ge 0$ gives our assertion.

Secondly, we assert that $D(x)$ is a \textit{decreasing} and \textit{positive} function on $x \in (\frac{3}{4}, t]$. By elementary algebra,
\begin{align*}
    D'(x)
    &\textstyle ~=~ \frac{y(x)}{(t - x)^{2}},\\
    y(x)
    &\textstyle ~\defeq~ (\frac{t}{b} + \frac{6x - 3t}{4x^{2}} - 1)
    ~-~ \big(a \cdot (x - \frac{3}{4}) \cdot \frac{t - x}{x} + \frac{6x - 3t}{4x^{2}} - 1\big) \cdot (\frac{3}{4})^{x}\\
    &\textstyle \phantom{~\defeq~} ~-~ \big((t - 2a x \cdot (t - x)) \cdot (1 - (\frac{3}{4})^{x}) + t\big) \cdot \frac{1}{b} \cdot (\frac{3}{4})^{x}.
\end{align*}
The function $y(x)$ is \textit{increasing} on the interval $x \in (\frac{3}{4}, t]$; by elementary algebra,
\begin{align*}
    y'(x)
    &\textstyle ~=~ \big(\frac{3}{2x^{2}} \cdot \big((\frac{4}{3})^{x} - (1 + ax)\big) + a^{2} \cdot (x - \frac{3}{4})\big) \cdot \frac{t - x}{x} \cdot (\frac{3}{4})^{x}\\
    &\textstyle \phantom{~=~} ~+~ \big(2 \cdot (1 - (\frac{3}{4})^{x}) \cdot (1 - a x) + a x\big) \cdot \frac{2a}{b} \cdot (t - x) \cdot (\frac{3}{4})^{x}\\
    &\textstyle ~\ge~ 0.
\end{align*}
Here the last step uses $t \in [\frac{7}{8} + \eps, 1]$,\quad
$a = \ln(\frac{4}{3}) \approx 0.2877$,\quad
and $(\frac{4}{3})^{x} \ge 1 + ax$.

The monotonicity of $y(x)$ together with $y(t) = 0$ (elementary algebra) implies $D'(x) \le 0$, for $x \in (\frac{3}{4}, t]$.
This in combination with $D(t) = 0$ (elementary algebra) gives our assertion.

Clearly, combining both assertions finishes the proof of \Cref{claim:three-MHR:MHR:h3}.
\end{proof}

The modified function $F_{3}^{t}$ is $(\frac{3}{4}, t]$-supported --- with different defining formulae on the sub-intervals $(\frac{3}{4}, s]$ and $(s, t]$ --- is \textit{continuous} at the division point $\lim_{x \to s^{+}} F_{3}^{t}(x) = F_{3}^{*t}(s) = F_{3}^{t}(s)$, has a \textit{positive} derivative function $f_{3}^{t}$ (as follows), takes values between $0 = F_{3}^{t}(\frac{3}{4}) \le F_{3}^{t}(t) \le 1$, and has a \textit{increasing} hazard rate function $h_{3}^{t}$ (as follows).
Given these, $F_{3}^{t}$ is a well-defined MHR CDF.
\begin{align*}
    f_{3}^{t}(x)
    & ~=~ F_{3}^{t}{}'(x) ~=~
    \begin{cases}
        f_{3}^{*t}(s) \cdot e^{-\frac{f_{3}^{*t}(s)}{1 - F_{3}^{*t}(s)} \cdot (x - s)}
         ~=~ f_{3}^{*t}(s) \cdot e^{-h_{3}^{*t}(s) \cdot (x - s)},
        & x \in (s, t]\\
        f_{3}^{*t}(x),
        & x \notin (s, t]
    \end{cases}, \\
    h_{3}^{t}(x)
    & ~=~ \tfrac{f_{3}^{t}(x)}{1 - F_{3}^{t}(x)} ~=~
    \begin{cases}
        \frac{f_{3}^{*t}(s) \cdot e^{-h_{3}^{*t}(s) \cdot (x - s)}}{(1 - F_{3}^{*t}(s)) \cdot e^{-h_{3}^{*t}(s) \cdot (x - s)}}
        ~=~ h_{3}^{*t}(s),
        & x \in (s, t]\\
        h_{3}^{*t}(x),
        & x \notin (s, t]
    \end{cases}.
\end{align*}

This finishes the proof of \Cref{lem:three-MHR:MHR}.
\end{proof}

Further, the following \Cref{lem:three-MHR:first-order,lem:three-MHR:revenue} will be useful for our lower-bound analysis in \Cref{subsec:three-MHR:analysis}.

\begin{lemma}[First-Order CDF's]
\label{lem:three-MHR:first-order}
$0 \le F^{*t}(x) - F^{t}(x) \le \eps$ for $x \in (s, t]$, while $F^{*t}(x) = F^{t}(x)$ for $x \notin (s, t]$.
\end{lemma}

\begin{proof}
Following the proof of \Cref{lem:three-MHR:MHR}, we have $h_{3}^{*t}(x) \ge h_{3}^{*t}(s) = h_{3}^{t}(x) \iff -\frac{f_{3}^{*t}(x)}{1 - F_{3}^{*t}(x)} \le -\frac{f_{3}^{t}(x)}{1 - F_{3}^{t}(x)} \iff \frac{\d}{\d x}\ln(1 - F_{3}^{*t}(x)) \le \frac{\d}{\d x}\ln(1 - F_{3}^{t}(x))$, for $x \in (s, t]$.
Together with the boundary condition $F_{3}^{*t}(s) = F_{3}^{t}(s)$, we deduce that $F_{3}^{*t}(x) - F_{3}^{t}(x) \ge 0 \implies F^{*t}(x) - F^{t}(x) = F_{1}^{*t}(x) \cdot F_{2}^{*t}(x) \cdot \big(F_{3}^{*t}(x) - F_{3}^{t}(x)\big) \ge 0$, for $x \in (s, t]$.
In addition, we have $f^{*t}(x) - f^{t}(x) \le f^{*t}(x) = \frac{3}{4x^{2}} \le 1$, for $x \in (s, t] \subseteq [\frac{7}{8}, 1]$, where the first step uses the modified first-order PDF $f^{t}$'s positivity. Together with the boundary condition $F^{*t}(s) = F^{t}(s)$, we deduce that $F^{*t}(x) - F^{t}(x) \le 1 \cdot (x - s) \le t - s = \eps$, for $x \in (s, t]$.

The second part ``$F^{*t}(x) = F^{t}(x)$ for $x \notin (s, t]$'' is obvious. This finishes the proof of \Cref{lem:three-MHR:first-order}.
\end{proof}

\begin{lemma}[Revenue Functions]
\label{lem:three-MHR:revenue}
$R^{t}(t) \ge \frac{3}{4} + \frac{1}{60}\eps$, while $R^{t}(x) \le \frac{3}{4}$ for $x \notin (s, t]$.
\end{lemma}

\begin{proof}
Note that
$t \in [\frac{7}{8} + \eps, 1]$,\quad
$s = t - \eps$,\quad
$F_{1}^{*t}(t) \cdot F_{2}^{*t}(t) = 1 - \frac{3}{4t}$,\quad
$F_{3}^{*t}(t) = 1$,\quad and\quad
$R^{*t}(t) = \frac{3}{4}$.
Using the defining formulae of $R^{t}(t)$, $R^{*t}(t)$, and $F_{3}^{t}(t)$, we can deduce ``$R^{t}(t) \ge \frac{3}{4} + \frac{1}{60}\eps$'' as follows:
\begin{align*}
    R^{t}(t) - R^{*t}(t)
    &\textstyle ~=~ t \cdot F_{1}^{*t}(t) \cdot F_{2}^{*t}(t) \cdot (F_{3}^{*t}(t) - F_{3}^{*t}(s)) \cdot e^{-h_{3}^{*t}(s) \cdot (t - s)}\\
    &\textstyle ~=~ (t - \frac{3}{4}) \cdot \big(\frac{F_{3}^{*t}(t) - F_{3}^{*t}(s)}{t - s} \cdot \eps\big) \cdot e^{-h_{3}^{*t}(t - \eps) \cdot \eps}\\
    &\textstyle ~\ge~ (t - \frac{3}{4}) \cdot \big(f_{3}^{*t}(1) \cdot \eps\big) \cdot e^{-h_{3}^{*t}(\frac{3}{4}) \cdot (t - \frac{3}{4})}\\
    &\textstyle ~\ge~ (t - \frac{3}{4}) \cdot (b \cdot \eps) \cdot e^{-9b \cdot (t - \frac{3}{4})}\\
    &\textstyle ~\ge~ \frac{1}{60}\eps.
\end{align*}
Here the third step uses the mean value theorem and \Cref{claim:three-MHR:revenue:1,claim:three-MHR:revenue:2}.
The fourth step uses $a = \ln(\frac{4}{3}) \approx 0.2877 \implies f_{3}^{*t}(1) = (2 - 3a) \cdot b \approx 1.1370 \cdot b \ge b$ and $h_{3}^{*t}(\frac{3}{4}) = \frac{16 / 9}{1 - (3 / 4)^{3 / 4}} \cdot b \approx 9.1604 \cdot b \ge 9b$.
And the last step uses $t \in [\frac{7}{8} + \eps, 1] \implies b \cdot (t - \frac{3}{4}) = t^{2} \cdot (1 - (\frac{3}{4})^{t}) \in [\frac{49 \cdot (1 - (3 / 4)^{7 / 8})}{64} \approx 0.1703,\ \frac{1}{4}] \subseteq [\frac{1}{6}, \frac{1}{4}]$, which implies that $b \cdot (t - \frac{3}{4}) \cdot e^{-9b \cdot (t - \frac{3}{4})} \ge \frac{1}{6}e^{-9 / 4} \approx 0.0176 \ge \frac{1}{60}$.

\begin{claim}
\label{claim:three-MHR:revenue:1}
$f_{3}^{*t}(x)$ is decreasing on $x \in [\frac{3}{4}, 1]$.
\Comment{Note that $[s, t] \subseteq [\frac{7}{8}, 1] \subseteq [\frac{3}{4}, 1]$.}
\end{claim}

\begin{proof}
Recall that $f_{3}^{*t}(x) = \frac{(6 - 4x) - (\frac{3}{4})^{x} \cdot (4a \cdot x^{2} - (3a + 4) \cdot x + 6)}{(1 - (\frac{3}{4})^{x})^{2}} \cdot \frac{b}{4x^{3}}$. By elementary algebra,
\begin{align*}
    f_{3}^{*t}{}'(x)
    &\textstyle ~=~ -\frac{b}{4x^{3} \cdot (1 - (\frac{3}{4})^{x})^{3}} \cdot y(x),\\
    y(x)
    &\textstyle ~\defeq~ 4a \cdot (\frac{3}{4})^{x} \cdot \underbrace{\big((3 - 2x) \cdot (1 - (\tfrac{3}{4})^{x}) - ax \cdot (x - \tfrac{3}{4}) \cdot (1 + (\tfrac{3}{4})^{x})\big)}_{\varheartsuit}
    ~+~ \underbrace{(\tfrac{18}{x} - 8) \cdot (1 - (\tfrac{3}{4})^{x})^{2}\big.}_{\ge 0}
\end{align*}
Given that $x \in [\frac{3}{4}, 1]$ and $a = \ln(\frac{4}{3}) \approx 0.2877$, we have $\varheartsuit \ge (3 - 2) \cdot (1 - (\frac{3}{4})^{3 / 4}) - a \cdot \frac{1}{4} \cdot (1 + \frac{3}{4}) \approx 0.0682 > 0$, which implies that $y(x) \ge 0 \implies f_{3}^{*t}{}'(x) \le 0$.
This finishes the proof of \Cref{claim:three-MHR:revenue:1}.
\end{proof}

\begin{claim}
\label{claim:three-MHR:revenue:2}
$y(x) \defeq x \cdot h_{3}^{*t}(t - x)$ is increasing on $x \in [0, t - \frac{3}{4}]$.
\Comment{Note that $t \in [\frac{7}{8} + \eps, 1] \implies \eps \in [0, t - \frac{3}{4}]$.}
\end{claim}

\begin{proof}
Recall that $h_{3}^{*t}(x) = \frac{f_{3}^{*t}(x)}{1 - F_{3}^{*t}(x)}$. By elementary algebra,
\begin{align*}
    \textstyle
    y'(x) ~=~ x \cdot h_{3}^{*t}(t - x) \cdot \frac{\d}{\d x} \big(\ln(x \cdot h_{3}^{*t}(t - x))\big).
\end{align*}
Let us substitute $z = (t - x) \in [\frac{3}{4}, t]$. Clearly, to show that $y(x)$ is \textit{increasing} on $x \in [0, t - \frac{3}{4}]$, it suffices to show that, under the same $t \in [\frac{7}{8}, 1]$, the function $g(z, t)$ is \textit{decreasing} on $z \in [\frac{3}{4}, t]$.
\begin{align*}
    g(z, t)
    &\textstyle ~\defeq~ (t - z) \cdot h_{3}^{*t}(z)
    ~=~ \frac{N(z, t)}{D(z, t)},\\
    N(z, t)
    &\textstyle ~\defeq~ \frac{6 - 4z}{z \cdot (4 - 3z)} - a \cdot \frac{4z - 3}{((\frac{4}{3})^{z} - 1) \cdot (4 - 3z)},\\
    D(z, t)
    &\textstyle ~\defeq~ \frac{(1 - (\frac{3}{4})^{z}) \cdot \frac{4z^{2}}{b} - (4z - 3)}{t - z} \cdot \frac{1}{4 - 3z}.
\end{align*}

Firstly, we assert that $N(z, t)$ is \textit{decreasing} and \textit{positive} $N(z, t) \ge N(1, t) = 2 - 3a \approx 1.1370$, even on the wider interval $z \in [\frac{3}{4}, 1]$. By rewriting $N(z, t) = N_{1}(z, t) - a \cdot \frac{N_{2}(z, t)}{N_{3}(z, t)}$, the monotonicity follows from a combination of three observations:\\
(i)~The function $N_{1}(z, t) \defeq \frac{6 - 4z}{z \cdot (4 - 3z)}$ is \textit{decreasing} $\frac{\partial}{\partial z} N_{1}(z, t) = -12 \cdot \frac{(1 - z) \cdot (2 - z)}{z^{2} \cdot (4 - 3z)^{2}} \le 0$.\\
(ii)~The function $N_{2}(z, t) \defeq 4z - 3$ is \textit{increasing} (obvious) and \textit{positive} (obvious).\\
(iii)~The function $N_{3}(z, t) \defeq ((\frac{4}{3})^{z} - 1) \cdot (4 - 3z)$ is \textit{decreasing} $\frac{\partial}{\partial z} N_{3}(z, t) = 3 \cdot (\frac{4}{3})^{x} \cdot ((\frac{3}{4})^{x} - a x - 1 + \frac{4}{3}a) \le 0$, given that $(\frac{3}{4})^{x} - a x - 1 + \frac{4}{3}a \le (\frac{3}{4})^{3 / 4} - \frac{3}{4}a - 1 + \frac{4}{3}a \approx -0.0263$, and \textit{positive} (obvious).

Secondly, we assert that $D(z, t)$ is \textit{increasing} and \textit{positive} $D(z, t) \ge D(\frac{3}{4}, t) = \frac{9}{7}t^{-2} \cdot \frac{1 - (3 / 4)^{3 / 4}}{1 - (3 / 4)^{t}} \ge 0$, even on the wider interval $z \in [\frac{3}{4}, 1]$. To see the monotonicity, by elementary algebra,
\begin{align*}
    \textstyle
    \frac{\partial}{\partial z}D(z, t)
    &\textstyle ~=~ \frac{1}{(t - z)^{2} \cdot (4 - 3x)^{2}} \cdot \frac{4}{b} \cdot K^{z}(t),\\
    K^{z}(t)
    &\textstyle ~\defeq~ \big(z^{2} \cdot (1 - (\frac{3}{4})^{z}) \cdot (4 - 3t) - t^{2} \cdot (1 - (\frac{3}{4})^{t}) \cdot (4 - 3z) \cdot \frac{4z - 3}{4t - 3}\big)\\
    &\textstyle \phantom{~\defeq~} ~+~ (t - z) \cdot z \cdot \big(a z \cdot (4 - 3z) \cdot (\frac{3}{4})^{z} + 8 \cdot (1 - (\frac{3}{4})^{z})\big)\\
    &\textstyle \phantom{~\defeq~} ~-~ (t - z) \cdot (1 - (\frac{3}{4})^{t}) \cdot \frac{7t^{2}}{4t - 3}.
\end{align*}
Clearly, it suffices to show $K^{z}(t) \ge 0$ for $z, t \in [\frac{3}{4}, 1]$. Below, given $z \in [\frac{3}{4}, 1]$, let us analyze this function. By elementary algebra, its first-order derivative $K^{z}{}'(t)$ and third-order derivative $K^{z}{}'''(t)$ are given by
\begin{align*}
    \textstyle
    K^{z}{}'(t)
    &\textstyle ~=~ (8z - 3z^{2}) \cdot \big((\frac{3}{4})^{t} - (\frac{3}{4})^{z}\big)
    ~+~ a \cdot \frac{4 -3z}{4z - 3} \cdot \big(\frac{z^{2}}{4z - 3} \cdot (\frac{3}{4})^{z} - \frac{t^{2}}{4t - 3} \cdot (\frac{3}{4})^{t}\big)\\
    &\textstyle \phantom{~=~} ~+~ (t - z) \cdot \big(\frac{-56t^{2} + 111t - 72 + 27z}{(4t-3)^{2}} \cdot \big(1 - (\frac{3}{4})^{t}\big) - \frac{7at^{2}}{4t-3}\cdot(\frac{3}{4})^{t}\big),\\
    \textstyle
    K^{z}{}'''(t)
    &\textstyle ~=~ -\frac{7a}{4} \cdot (\frac{3}{4})^{t} \cdot (a^{2} t^{2} - 6a t + 6) - 162 \cdot \frac{(4u - 3)^{2}}{(4t - 3)^{4}} \cdot \big(1 - (\frac{3}{4})^{t} \cdot L(t)\big),\\
    \textstyle
    L(t)
    &\textstyle ~\defeq~ 1 + \frac{3}{4} \cdot (\frac{4}{3}t - 1) \cdot a + \frac{t \cdot (8t^{2} - 14t + 9)}{12} \cdot (\frac{4}{3}t - 1)^{2} \cdot a^{2} + \frac{t^{2}}{8} \cdot (\frac{4}{3}t - 1)^{3} \cdot a^{3}.
\end{align*}
Also, we have $\big(a = \ln(\frac{4}{3}) \approx 0.2877\big) \land \big(t \in [\frac{3}{4}, 1]\big) \implies a^{2} t^{2} - 6a t + 6 \ge 0$ and ``$L(t)$ is increasing on $t \in [\frac{3}{4}, 1]$'' $\implies L(t) \le L(1) = 1 + \frac{a}{4} + \frac{a^{2}}{36} + \frac{a^{3}}{216} \approx 1.0743 \le (\frac{4}{3})^{3 / 4} \approx 1.2408 \le (\frac{4}{3})^{t}$. Hence, on the interval $t \in [\frac{3}{4}, 1]$, the third-order derivative is \textit{negative} $K^{z}{}'''(t) \le 0$ and the first-order derivative $K^{z}{}'(t)$ is \textit{concave}.

At the point $t = z$, the second-order derivative $K^{z}{}''(t)$ evaluates to $K^{z}{}''(z) = M(z) \ge 0$, where
\begin{align*}
    \textstyle
    M(z)
    &\textstyle ~\defeq~ \underbrace{\textstyle \frac{24 - 14z}{4z - 3} \cdot (1 - (\frac{3}{4})^{z})}_{\heartsuit}
    ~+~ \underbrace{\textstyle \frac{a^{2} \cdot (3 / 4)^{z}}{(4z - 3)^{2}} \cdot 9z^{3} \cdot (\frac{4}{3} - z) \cdot (\frac{8}{3} - z)}_{\ge 0}\\
    &\textstyle \phantom{~\defeq~} ~+~ \frac{a \cdot (3 / 4)^{z}}{(4z - 3)^{3}} \cdot \big((1 - z) \cdot \underbrace{(-156z^{4} + 702z^{3} - 598z^{2} + 218z + 2)}_{\vardiamondsuit} - 2\big)\\
    &\textstyle ~\ge~ \frac{5}{2} + \frac{a \cdot (3 / 4)^{z}}{(4z - 3)^{3}} \cdot \big((1 - z) \cdot 168 \cdot (3z - 2) - 2\big)\\
    &\textstyle ~\ge~ \frac{5}{2} + \frac{1}{(4z - 3)^{3}} \cdot 36 \cdot (1 - z) \cdot (3z - 2) - \frac{1}{(4z - 3)^{3}} \cdot \frac{1}{2}\\
    &\textstyle ~=~ \frac{1}{(4z - 3)^{3}} \cdot \big(\frac{7}{4} + \underbrace{\textstyle 160 \cdot (z - \frac{3}{4}) \cdot (\frac{9}{8} - z) \cdot (\frac{21}{20} - z)}_{\ge 0}\big)\\
    &\textstyle ~\ge~ 0.
\end{align*}
Here the second step uses $\heartsuit \ge \frac{24 - 14z}{4z - 3} \cdot \frac{z}{4} \ge \frac{5}{2}$ and $\vardiamondsuit = 168 \cdot (3z - 2) + 26 \cdot (1 - z)^{2} \cdot (-6z^{2} + 15z + 13) \ge 168 \cdot (3z - 2)$ for $z \in [\frac{3}{4}, 1]$, both of which can be verified via elementary algebra.
And the third step uses $a \cdot (3 / 4)^{z} \cdot 168 \ge 126a \approx 36.2479 \ge 36$ and $2a \cdot (3 / 4)^{z} \le 2a \cdot (3 / 4)^{3 / 4} \approx 0.4637 \le \frac{1}{2}$ for $z \in [\frac{3}{4}, 1]$.

At the point $t = z$, the first-order derivative $K^{z}{}'(t)$ equals zero $K^{z}{}'(z) = 0$ (obvious). Together with its \textit{concavity} and that $K^{z}{}''(z) = M(z) \ge 0$, this means that $K^{z}(t)$ is \textit{decreasing} on the interval $t \in [\frac{3}{4}, z]$ and is \textit{``either increasing or first-increasing-then-decreasing''} on the interval $t \in [z, 1]$; in any case, we always have
\begin{align*}
    \textstyle
    \min_{t \in [\frac{3}{4}, 1]} K^{z}(t)
    ~=~ \min\big(K^{z}(z),\ K^{z}(1)\big).
\end{align*}

The function $K^{z}(t)$ equals zero $K^{z}(z) = 0$ at the point $t = z$ (obvious). Accordingly, to show $K^{z}(t) \ge 0$ for $t \in [\frac{3}{4}, 1]$, it remains to show $K^{z}(1) \ge 0$, as follows.
\begin{align*}
    K^{z}(1)
    &\textstyle ~=~ (\frac{3}{4})^{z} \cdot z \cdot((1 - z) \cdot a z \cdot (4 - 3z) - (8 - 7z)) - \frac{(16z^{2} - 14z - 5)}{4}\\
    &\textstyle ~=~ (\frac{3}{4})^{z} \cdot (1 - z) \cdot \big((a - \frac{1 - (3 / 4)^{1 - z}}{1 - z}) + ((\frac{3}{4})^{1 - z} - 1) \cdot \frac{16z + 2}{3} + (1 - z) \cdot \big(\frac{5}{3} + a \cdot(3z^{2} - z - 1)\big)\big)\\
    &\textstyle ~\ge~ (\frac{3}{4})^{z} \cdot (1 - z) \cdot \big(0 - a \cdot (1 - z) \cdot \frac{16z + 2}{3} + (1 - z) \cdot (\frac{5}{3} + a \cdot (3z^{2} - z - 1))\big)\\
    &\textstyle ~=~ (\frac{3}{4})^{z}\cdot(1-z)^{2}\cdot((\frac{5}{3}-5a)+\frac{a}{3}(1-z)+3a(1-z)^{2})\\
    &\textstyle ~\ge~ 0.
\end{align*}
Here the second step rearranges the equation.
The third step uses $\frac{1 - (3 / 4)^{1 - z}}{1 - z} \le a = \ln(\frac{4}{3}) \approx 0.2877$ for $z \in [\frac{3}{4}, 1]$. And the fourth step rearranges the equation.

This finishes the proof of \Cref{claim:three-MHR:revenue:2}.
\end{proof}

The second part ``$R^{t}(x) = R^{*t}(x) \le \frac{3}{4}$ for $x \notin (s, t]$'' is obvious. This finishes the proof of \Cref{lem:three-MHR:revenue}.
\end{proof}

To obtain the desired lower bound $\Omega(\eps^{-3})$, we choose a sequence of \textit{disjoint} modification intervals\footnote{Since $K = \lfloor \frac{1}{8}\eps^{-1} \rfloor$, all parameters $\{t^{i}\}_{i \in [K]}$ belong to the interval $[\frac{7}{8} + \eps, 1]$, satisfying the premises of \Cref{lem:three-MHR:MHR,lem:three-MHR:revenue}.} $(s^{i}, t^{i}] \defeq (\frac{7}{8} + i \eps - \eps, \frac{7}{8} + i \eps]$ and, thus, obtain our hard instances $\bF^{i} \defeq F_{1}^{*t^{i}} \otimes F_{2}^{*t^{i}} \otimes F_{3}^{t^{i}}$ for $i \in [K]$.

\subsection{Lower Bound Analysis}
\label{subsec:three-MHR:analysis}

Consider a specific pricing algorithm $\+A$ that performs well on all hard instances $\{\bF^{i}\}_{i \in [K]}$: in any possibility $i \in [K]$, it always outputs a $\frac{1}{70}\eps$-approximately optimal price $p^{\+A}$ with probability $\ge \frac{2}{3}$:
\begin{align*}
    \textstyle
    R^{i}(p^{\+A})
    ~\ge~ \max_{p \in [0, 1]} R^{i}(p) - \frac{1}{70}\eps.
\end{align*}
(In contrast, no performance guarantee for the base instance $\bF^{*}$ is needed.)
Based on this, we can develop a ``pricing-to-identification'' reduction for all base/hard instances $\{\bF^{*}\} \cup \{\bF^{i}\}_{i \in [K]}$, i.e., another identification algorithm $\+B^{\+A}$ with \textit{exactly the same number of pricing queries}:
\begin{itemize}
    \item Run $\+A$ on an unknown instance $\bF$ (promised to be one of $\{\bF^{*}\} \cup \{\bF^{i}\}_{i \in [K]}$), getting a price $p^{\+A}$.
    
    \item If $p^{\+A} \in (s^{i}, t^{i}]$ for some $i \in [K]$, output $\bF^{i}$;
    
    \item Otherwise, output $\bF^{*}$.
\end{itemize}
Clearly, $\+B^{\+A}$ can identify all hard instances $\{\bF^{i}\}_{i \in [K]}$:\footnote{I.e., this directly follows from a combination of \Cref{lem:three-MHR:revenue}, that the modification intervals $(s^{i}, t^{i}]$ are disjoint, and $\+A$'s performance guarantees.}
in any possibility $i \in [K]$, it always successfully outputs $\bF^{i}$ with probability $\ge \frac{2}{3}$.
(Again, no performance guarantee for the base instance $\bF^{*}$ is needed.)

In the rest of this section, we consider a specific identification algorithm $\+B$; denote by $T$ the number of pricing queries it makes and, in particular, $T^{i}\ignore{ = \big|\big\{t \in [T]: p_t \in (s^{i}, t^{i}]\big\}\big|}$ for $i \in [K]$ the number of pricing queries it makes within the index-$i$ modification interval $(s^{i}, t^{i}]$. Also, denote by $\bb{P}^{*}[\cdot]$ or $\bb{P}^{i}[\cdot]$ the probabilities in each possibility $i \in [K]$; likewise for the expectations $\bb{E}^{*}[\cdot]$ or $\bb{E}^{i}[\cdot]$.

The following \Cref{lem:three-MHR:identify} lower-bounds the query complexity of an identification algorithm $\+B$.

\begin{lemma}[Identification Lower Bounds]
\label{lem:three-MHR:identify}
To identify hard instances $\{\bF^{i}\}_{i \in [K]}$ each with probability $\ge \frac{2}{3}$, an identification algorithm $\+B$ makes at least $\bb{E}^{*}[T] = \Omega(\eps^{-3})$ many pricing queries on the base instance $\bF^{*}$ (in expectation over the randomness of both $\+B$ itself and $\bF^{*}$).
\end{lemma}

\begin{proof}
Consider the base instance $\bF^{*}$ and a specific hard instance $\bF^{i}$.
As mentioned, only pricing queries within the corresponding modification interval $(s^{i}, t^{i}]$ can help identify $\bF^{i}$.

Recall that $\DKL(p, q) = p \ln(\frac{p}{q}) + (1 - p) \ln(\frac{1 - p}{1 - q})$ denotes the KL divergence between two Bernoulli distributions with parameters $p, q \in [0, 1]$.
For $x \in (s^{i}, t^{i}] \subseteq [\frac{7}{8}, 1]$, we have $0 \le F^{*}(x) - F^{i}(x) \le \eps \le \frac{1}{48}$ (\Cref{lem:three-MHR:first-order}) and $F^{*}(x) = 1 - \frac{3}{4x} \in [\frac{1}{7}, \frac{1}{4}]$, so \Cref{cla:DKL-UB} is applicable and gives
\begin{align*}
    \textstyle
    \DKL\big(F^{*}(x),\ F^{i}(x)\big)
    ~\le~ 3\eps^{2}.
\end{align*}
Then, regarding the event $\@E^{i} \defeq \{\text{$\+B$ outputs $\bF^{i}$}\}$, we know from \Cref{lem:kl-bound} that
\begin{align*}
    \textstyle
    \DKL\big(\bb{P}^{*}[\@E^{i}],\ \bb{P}^{i}[\@E^{i}]\big)
    ~\le~ \DKL\big(F^{*}(x),\ F^{i}(x)\big) \cdot \bb{E}^{*}[T^{i}]
    ~\le~ 3\eps^{2} \cdot \bb{E}^{*}[T^{i}].
\end{align*}
By enumerating all $i \in [K]$, we can upper-bound the sum $\sum_{i \in [K]} \DKL(\bb{P}^{*}[\@E^{i}],\ \bb{P}^{i}[\@E^{i}])$ as follows:
\begin{align*}
    \textstyle
    \sum_{i \in [K]} \DKL\big(\bb{P}^{*}[\@E^{i}],\ \bb{P}^{i}[\@E^{i}]\big)
    ~\le~ \sum_{i \in [K]} 3\eps^{2} \cdot \bb{E}^{*}[T^{i}]
    ~\le~ 3\eps^{2} \cdot \bb{E}^{*}[T].
\end{align*}
Here the last step uses the linearity of expectations and that $\sum_{i \in [K]} T_{i} \le T$ (almost surely over all possible randomness).

Moreover, since the KL divergence $\DKL(p, q)$ is a convex function (\Cref{cla:DKL-convex}), using Jensen's inequality (\Cref{cla:Jensen}), we can lower-bound the sum $\sum_{i \in [K]} \DKL(\bb{P}^{*}[\@E^{i}],\ \bb{P}^{i}[\@E^{i}])$ as follows:
\begin{align*}
    \textstyle
    \sum_{i \in [K]} \DKL\big(\bb{P}^{*}[\@E^{i}],\ \bb{P}^{i}[\@E^{i}]\big)
    &\textstyle ~\ge~ K \cdot \DKL\Big(\frac{\sum_{i \in [K]} \bb{P}^{*}[\@E^{i}]}{K},\ \frac{\sum_{i \in [K]} \bb{P}^{i}[\@E^{i}]}{K}\Big)\\
    &\textstyle ~\ge~ K \cdot \DKL(\frac{1}{6}, \frac{2}{3})\\
    &\textstyle ~\ge~ \frac{1}{2}K.
\end{align*}
Here the second step uses ``$\{\@E^{i}\}_{i \in [K]}$ are disjoint'' $\implies \frac{\sum_{i \in [K]} \bb{P}^{*}[\@E^{i}]}{K} \le \frac{1}{K}\le \frac{1}{6}$ and the premise of the lemma ``$\bb{P}^{i}[\@E^{i}] \ge \frac{2}{3}$ for $i \in [K]$''.
And the last step uses $\DKL(\frac{1}{6}, \frac{2}{3}) = \frac{5}{6}\ln 5 - \frac{7}{6}\ln 2 \approx 0.5325$.

Combining the above two equations directly gives $\bb{E}^{*}[T] \ge \frac{1}{6}\eps^{-2} \cdot K \ge \frac{5}{288}\eps^{-3}$, where the last step uses $\eps \in (0, \frac{1}{48}) \implies K = \lfloor \frac{1}{8}\eps^{-1} \rfloor \ge \frac{5}{48}\eps^{-1}$. This finishes the proof of \Cref{lem:three-MHR:identify}.
\end{proof}

Finally, we can translate the query complexity lower bound of an identification algorithm (\Cref{lem:three-MHR:identify}) into that of a pricing algorithm (\Cref{thm:three-MHR}).

\begin{proof}[Proof of \Cref{thm:three-MHR}]
Incorporate the ``pricing-to-identification'' reduction into \Cref{lem:three-MHR:identify}: if a pricing algorithm $\+A$ always outputs a $\frac{1}{70}\eps$-approximately optimal price $p^{\+A}$ with probability $\ge \frac{2}{3}$, then it makes at least $\Omega(\eps^{-3})$ many pricing queries on the base instance $\bF^{*}$.

Scaling the parameter $\eps \in (0, \frac{1}{48})$ by a factor of $\frac{1}{70}$ finishes the proof of \Cref{thm:three-MHR}.
\end{proof}

\begin{flushleft}
\bibliographystyle{alphaurl}
\bibliography{main}
\end{flushleft}

\appendix

\section{\texorpdfstring{$\Omega(\eps^{-5 / 2})$ Lower Bound for Two MHR Distributions}{}}
\label{sec:two-MHR}

In this appendix, we investigate the query complexity of {\UniformPricing} in the setting with \textit{two MHR distributions}. Specifically, we will establish the following \Cref{thm:two-MHR}.

\begin{theorem}
\label{thm:two-MHR}
For two MHR distributions, the query complexity of {\UniformPricing} is $\Omega(\eps^{-5 / 2})$.
\end{theorem}

In the remainder of \Cref{thm:two-MHR}, we consider a sufficiently small $\eps \in (0, \frac{1}{100})$ and a sufficiently large $K \defeq \lfloor \frac{1}{16}\eps^{-1} \rfloor \ge 6$ throughout.
The overall approach is symmetric to \Cref{sec:regular,sec:three-MHR}; for brevity, we only present counterparts of the lower-bound construction (\Cref{subsec:two-MHR-construction}) and the technical lemmas (\Cref{subsec:two-MHR-lemma}). Without ambiguity, we often reload the notations introduced in \Cref{subsec:regular:construction,subsec:three-MHR:construction}.

\subsection{Lower Bound Construction}
\label{subsec:two-MHR-construction}

We define our base instance $\bF^{*} \defeq F_{1}^{*} \otimes F_{2}^{*}$ as follows.
(We will verify the MHR condition later in \Cref{lem:two-MHR:MHR}.)
\begin{align*}
    F_{1}^{*}(x)
    &\textstyle ~\defeq~ 1 - (\frac{3}{4})^{x},
    &&\hspace{-4.87cm}\textstyle x \in [0, 1],\\
    F_{2}^{*}(x)
    & ~\defeq~
    \begin{cases}
        0,
        &\hspace{1.3cm} x \in [0, \frac{3}{4}]\\
        \frac{1 - \frac{3}{4x}}{1 - (\frac{3}{4})^{x}},
        &\hspace{1.3cm} x \in (\frac{3}{4}, 1]
    \end{cases}.
\end{align*}
In regard to {\UniformPricing}, the first-order CDF $F^{*}(x) \defeq F_{1}^{*}(x) \cdot F_{2}^{*}(x)$ and the revenue function $R^{*}(x) \defeq x \cdot (1 - F^{*}(x))$ are given as follows:
\begin{align*}
    F^{*}(x)
    & ~=~
    \begin{cases} 
        0, & x \in [0, \frac{3}{4}]\\
        \frac{x - \frac{3}{4}}{x}, & x \in (\frac{3}{4}, 1]
    \end{cases},\\
    R^{*}(x)
    & ~=~
    \begin{cases} 
        x, & x \in [0, \frac{3}{4}]\\
        \frac{3}{4}, & x \in (\frac{3}{4}, 1]
    \end{cases}.
\end{align*}

Consider two parameters $t \in [\frac{7}{8} + \eps, \frac{15}{16}]$ and $s = s(t) \defeq t - \eps$.
The \textit{cumulative hazard rate function} $H_{2}^{*}(x)$ is \textit{convex} over $x \in [0, 1]$, because its derivative $H_{2}^{*}{}'(x) \equiv h_{2}^{*}(x)$ is exactly the second CDF $F_{2}^{*}$'s \textit{increasing} hazard rate function --- we will prove this monotonicity in \Cref{lem:two-MHR:MHR}.
\begin{align*}
    H_{2}^{*}(x) ~\defeq~ -\ln(1 - F_{2}^{*}(x)).
\end{align*}
Hence, regarding the tangent lines of $H_{2}^{*}(x)$ at $x = s$ and $x = t$, respectively, the following equation has a unique solution $m = m(t) \in (s, t]$.
\begin{align*}
    & h_{2}^{*}(s) \cdot (m - s) + H_{2}^{*}(s)
    ~=~ h_{2}^{*}(t) \cdot (m - t) + H_{2}^{*}(t).
\end{align*}

To establish the desired lower bound $\Omega(\eps^{-5 / 2})$, we modify the second CDF $F_{2}^{*}$ into another parametric CDF $F_{2}^{t}$ on the interval $(s, t] = (t - \eps, t]$, as follows. (In contrast, the first CDF $F_{1}^{*}$ keeps the same.)
\begin{align*}
    F_{2}^{t}(x) ~\defeq~
    \begin{cases}
        1 - (1 - F_{2}^{*}(s)) \cdot e^{-\frac{f_{2}^{*}(s)}{1 - F_{2}^{*}(s)} \cdot (x - s)}
        ~=~ 1 - (1 - F_{2}^{*}(s)) \cdot e^{-h_{2}^{*}(s) \cdot (x - s)}, & x \in (s, m]\\
        1 - (1 - F_{2}^{*}(t)) \cdot e^{-\frac{f_{2}^{*}(t)}{1 - F_{2}^{*}(t)} \cdot (x - t)}
        ~=~ 1 - (1 - F_{2}^{*}(t)) \cdot e^{-h_{2}^{*}(t) \cdot (x - t)}, & x \in (m, t]\\
        F_{2}^{*}(x), & x \notin (s, t]
    \end{cases}.
\end{align*}
Only pricing queries within the modification interval $(s, t]$ can help identify $F_{2}^{t}$.
The modified first-order CDF $F^{t}(x) \defeq F_{1}^{*}(x) \cdot F_{2}^{t}(x)$ and the modified revenue function $R^{t}(x) \defeq x \cdot (1 - F^{t}(x))$ follow accordingly.

\begin{remark}
The modified parametric CDF $F_{2}^{t}$ is defined such that \textit{``over the interval $x \in (s, t]$, the corresponding hazard rate function $h_{2}^{t}(x)$ is two-piecewise constant, i.e., $h_{2}^{t}(x) = h_{2}^{*}(s)$ for $x \in [s, m)$ and $h_{2}^{t}(x) = h_{2}^{*}(t)$ for $x \in [m, t)$''}; below, the proof of \Cref{lem:two-MHR:MHR} will show this explicitly.
This induces a two-piecewise ODE, and solving it under the boundary conditions $\lim_{x \to s^{+}} F_{2}^{t}(x) = F_{2}^{*}(s)$ and $F_{2}^{t}(t) = F_{2}^{*}(t)$ (so as to preserve the continuity at $x = s$ and $x = t$) gives the above defining formula.
\end{remark}

To obtain the desired lower bound $\Omega(\eps^{-5 / 2})$, we choose a sequence of \textit{disjoint} modification intervals\footnote{Since $K = \lfloor \frac{1}{16}\eps^{-1} \rfloor$, all parameters $\{t^{i}\}_{i \in [K]}$ belong to the interval $[\frac{7}{8} + \eps, \frac{15}{16}]$, satisfying the premises of \Cref{lem:two-MHR:MHR,lem:two-MHR:revenue}.} $(s^{i}, t^{i}] \defeq (\frac{7}{8} + i \eps - \eps, \frac{7}{8} + i \eps]$ and, thus, obtain our hard instances $\bF^{i} \defeq F_{1}^{*} \otimes F_{2}^{t^{i}}$ for $i \in [K]$.

\subsection{Counterpart Lemmas}
\label{subsec:two-MHR-lemma}

Below, we present counterparts of the technical lemmas.

Firstly, the following \Cref{lem:two-MHR:MHR,lem:two-MHR:first-order,lem:two-MHR:revenue} are counterparts of \Cref{lem:three-MHR:MHR,lem:three-MHR:first-order,lem:three-MHR:revenue}, respectively.

\begin{lemma}[MHR]
\label{lem:two-MHR:MHR}
Given any $t \in [\frac{7}{8} + \eps, \frac{15}{16}]$, all of $F_{1}^{*}$, $F_{2}^{*}$, and $F_{2}^{t}$ are well-defined MHR CDF's.
\end{lemma}

\begin{proof}
The first function $F_{1}^{*}(x) = 1 - (\frac{3}{4})^{x}$ for $x \in [0, 1]$ is clearly a well-defined MHR CDF.

The second function $F_{2}^{*}$ is $(\frac{3}{4}, 1]$-supported, is \textit{continuous} at the two endpoints $\lim_{x \to (\frac{3}{4})^{+}} F_{2}^{*}(x) = 0 = F_{2}^{*}(\frac{3}{4})$ and $F_{2}^{*}(1) = 1$, has a \textit{positive} derivative function $f_{2}^{*}$ (which has already been shown in the proof of \Cref{claim:three-MHR:MHR:f2}), and has a \textit{increasing} hazard rate function $h_{2}^{*}$ (which has already been shown in the proof of \Cref{claim:three-MHR:MHR:h2}).
Given these, $F_{2}^{*}$ is a well-defined MHR CDF.

The modified function $F_{2}^{t}$ is $(\frac{3}{4}, 1]$-supported --- with different defining formulae on $(s, m]$, $(m, t]$, and elsewhere $x \notin (s, t]$ --- is \textit{continuous} at all the division points $s \le m \le t$,\footnote{I.e., we have $\lim_{x \to s^{+}} F_{2}^{t}(x) = F_{2}^{*t}(s) = F_{2}^{t}(s)$,\quad $\lim_{x \to t^{+}} F_{2}^{t}(t) = F_{2}^{*t}(t) = F_{2}^{t}(t)$,\quad and (given the defining equation for $m$) $\lim_{x \to m^{+}} F_{2}^{t}(x) = 1 - e^{-h_{2}^{*}(t) \cdot (m - t) - H_{2}^{*}(t)} = 1 - e^{-h_{2}^{*}(s) \cdot (m - s) - H_{2}^{*}(s)} = F_{2}^{t}(m)$.} has a \textit{positive} derivative function $f_{2}^{t}$ (as follows), takes values between $0 = F_{2}^{t}(\frac{3}{4}) \le F_{2}^{t}(t) \le 1$, and has a \textit{increasing} hazard rate function $h_{2}^{t}$ (as follows).
Given these, $F_{2}^{t}$ is a well-defined MHR CDF.
\begin{align*}
    f_{2}^{t}(x)
    & ~=~ F_{2}^{t}{}'(x) ~=~
    \begin{cases}
        f_{2}^{*}(s) \cdot e^{-h_{2}^{*}(s) \cdot (x - s)}, & x \in (s, m]\\
        f_{2}^{*}(t) \cdot e^{-h_{2}^{*}(t) \cdot (x - t)}, & x \in (m, t]\\
        f_{2}^{*}(x), & x \notin (s, t]
    \end{cases}, \\
    h_{2}^{t}(x)
    & ~=~ \tfrac{f_{2}^{t}(x)}{1 - F_{2}^{t}(x)} ~=~
    \begin{cases}
        h_{2}^{*}(s), & x \in (s, m]\\
        h_{2}^{*}(t), & x \in (m, t]\\
        h_{2}^{*}(x), & x \notin (s, t]
    \end{cases}.
\end{align*}

This finishes the proof of \Cref{lem:two-MHR:MHR}.
\end{proof}

\begin{lemma}[First-Order CDF's]
\label{lem:two-MHR:first-order}
{$0 \le F^{*}(x) - F^{t}(x) \le 20\eps^{2}$} for $x \in (s, t]$, while $F^{*}(x) = F^{t}(x)$ for $x \notin (s, t]$.
\end{lemma}

\begin{proof}
\Cref{lem:two-MHR:MHR} implies that the \textit{cumulative hazard rate function} $H_{2}^{*}(x) = -\ln(1 - F_{2}^{*}(x))$ is \textit{convex} over $x \in [0, 1]$. On the modification interval $x \in (s, t]$, the modified counterpart $H_{2}^{*}(x)$ is a two-piecewise linear function, (essentially) consisting of the tangent lines of $H_{2}^{*}(x)$ at $x = s$ and $x = t$.
\begin{align*}
    H_{2}^{t}(x)
    ~\defeq~ -\ln(1 - F_{2}^{t}(x))
    ~=~ 
    \begin{cases} 
        H_{2}^{*}(s) + h_{2}^{*}(s) \cdot (x - s), & x \in (s, m]\\
        H_{2}^{*}(t) + h_{2}^{*}(t) \cdot (x - t), & x \in (m, t]
    \end{cases}.
\end{align*}
Given the convexity of $H_{2}^{*}(x)$, we have $H_{2}^{t}(x) \le H_{2}^{*}(x) \implies F_{2}^{t}(x) \le F_{2}^{*}(t) \implies F^{t}(x) \le F^{*}(x)$, for $x\in (s,t]$.

The following \Cref{claim:two-MHR:first-order:1} will be useful for our later proofs.

\begin{claim}
\label{claim:two-MHR:first-order:1}
The second-order derivative $H_{2}^{*}{}''(x)$ is increasing on $x \in [\frac{7}{8}, \frac{15}{16}]$.
\end{claim}

\begin{proof}
Recall that $a = \ln(\frac{4}{3}) \approx 0.2877$. By elementary algebra, we can formulate the third-order derivative
\begin{align*}
    H_{2}^{*}{}'''(x)
    &\textstyle ~=~ 16 \cdot (\frac{3}{4})^{2x} \cdot \frac{1 - a x}{(3 - 4x \cdot (\frac{3}{4})^{x})^{3}} \cdot \underbrace{\textstyle \big(8 \cdot (\frac{3}{4})^{x} \cdot (1 - ax)^{2} - 3a \cdot (3 - 4x \cdot (\frac{3}{4})^{x}) \cdot (2 - ax)\big)}_{\vardiamondsuit}\\
    &\textstyle \phantom{~=~} ~+~ \underbrace{\textstyle a^{3} \cdot (\frac{3}{4})^{x} \cdot \frac{1 + (\frac{3}{4})^{x}}{(1 - (\frac{3}{4})^{x})^{3}}
    ~+~ 4a^{2} \cdot (\frac{3}{4})^{x} \cdot \frac{3 - a x}{3 - 4x \cdot (\frac{3}{4})^{x}}
    ~+~ \frac{2}{x^{3}}}_{\ge 0}
\end{align*}
It is easy to show that $\vardiamondsuit \ge 0$ for $x \in [\frac{7}{8}, \frac{15}{16}]$, as follows:
\begin{align*}
    \textstyle
    \vardiamondsuit
    ~\ge~ 8 \cdot (\frac{3}{4})^{15 / 16} \cdot (1 - a \cdot \frac{15}{16})^{2} - 3a \cdot (3 - 4 \cdot \frac{7}{8} \cdot (\frac{3}{4})^{15 / 16}) \cdot (2 - a \cdot \frac{7}{8})
    ~\approx~ 2.7641.
\end{align*}
Then, it is easy to see the nonnegativity of the third-order derivative $H_{2}^{*}{}'''(x) \ge 0$ and thus the monotonicity of the second-order derivative $H_{2}^{*}{}''(x)$.
This finishes the proof of \Cref{claim:two-MHR:first-order:1}.
\end{proof}

By Taylor's theorem, (i)~for $x \in [s, m]$, there exists $\lambda = \lambda(x) \in [s, x]$ such that
\begin{align*}
    H_{2}^{*}(x)
    ~=~ H_{2}^{*}(s)
    + h_{2}^{*}(s) \cdot (x - s)
    + \tfrac{1}{2}H_{2}^{*}{}''(\lambda) \cdot (x - s)^{2}
    ~=~ H_{2}^{t}(x) + \tfrac{1}{2}H_{2}^{*}{}''(\lambda) \cdot (x - s)^{2},
\end{align*}
and (ii)~for $x \in [m, t]$, there exists $\mu = \mu(x) \in [x, t]$ such that
\begin{align*}
    H_{2}^{*}(x)
    ~=~ H_{2}^{*}(t)
    + h_{2}^{*}(t) \cdot (x - t)
    + \tfrac{1}{2}H_{2}^{*}{}''(\mu) \cdot (x - t)^{2}
    ~=~ H_{2}^{t}(x) + \tfrac{1}{2}H_{2}^{*}{}''(\mu) \cdot (x - t)^{2}.
\end{align*}
Together with $H_{2}^{*}(x) = -\ln(1 - F_{2}^{*}(x))$ and $H_{2}^{t}(x) = -\ln(1 - F_{2}^{t}(x))$, we can deduce that
\begin{align*}
    F^{*}(x) - F^{t}(x)
    & ~=~ F_{1}^{*}(x) \cdot \big(F_{2}^{*}(x) - F_{2}^{t}(x)\big)\\
    & ~=~
    \begin{cases}
        F_{1}^{*}(x) \cdot (1 - F_{2}^{t}(x)) \cdot \big(1 - e^{-\frac{1}{2}H_{2}^{*}{}''(\lambda) \cdot (x - s)^{2}}\big), & x \in [s, m]\\
        F_{1}^{*}(x) \cdot (1 - F_{2}^{t}(x)) \cdot \big(1 - e^{-\frac{1}{2}H_{2}^{*}{}''(\mu) \cdot (x - t)^{2}}\big), & x \in [m, t]
    \end{cases}.
\end{align*}

For $x \in (s, t] \subseteq [\frac{7}{8}, \frac{15}{16}]$, we deduce that $F_{1}^{*}(x) \cdot (1 - F_{2}^{t}(x)) \le F_{1}^{*}(\frac{15}{16}) \cdot (1 - F_{2}^{t}(\frac{7}{8})) = F_{1}^{*}(\frac{15}{16}) \cdot (1 - F_{2}^{*}(\frac{7}{8})) \ignore{= (1 - (\frac{3}{4})^{15 / 16}) \cdot \frac{1 - \frac{3}{4 \cdot \frac{7}{8}}}{1 - (\frac{3}{4})^{\frac{7}{8}}}} \approx 0.1518 < \frac{7}{45}$.
Together with the facts that $1 - e^{-y} \le y$, that $\max\big(H_{2}^{*}{}''(\lambda),\ H_{2}^{*}{}''(\mu)\big) \le H_{2}^{*}{}''(\frac{15}{16}) \approx 253.7455 < 254$ (\Cref{claim:two-MHR:first-order:1}), and that $\max\big((x - s)^{2},\ (x - t)^{2}\big) \le (t - s)^{2} = \eps^{2}$, we can conclude with
\begin{align*}
    \textstyle
    F^{*}(x) - F^{t}(x)
    ~\le~ \frac{7}{45} \cdot \frac{1}{2} \cdot 254 \cdot \eps^{2}
    ~<~ 20\eps^{2}.
\end{align*}

This finishes the proof of \Cref{lem:two-MHR:first-order}.
\end{proof}

\begin{lemma}[Revenue Functions]
\label{lem:two-MHR:revenue}
{$R^{t}(m) \ge \frac{3}{4} + \eps^{2}$}, while $R^{t}(x) \le \frac{3}{4}$ for $x \notin (s, t]$.
\end{lemma}

\begin{proof}
Recall that $R^{t}(m) = m \cdot (1 - F^{t}(m))$ and $R^{*}(m) = m \cdot (1 - F^{*}(m))$.
Following the proof of \Cref{lem:two-MHR:first-order}, there exist $\lambda = \lambda(m) \in [s, m]$ and $\mu = \mu(m) \in [m, t]$ such that
\begin{align*}
    R^{t}(m) - R^{*}(m)
    & ~=~ m \cdot F_{1}^{*}(m) \cdot (1 - F_{2}^{t}(m)) \cdot \big(1 - e^{-\frac{1}{2}H_{2}^{*}{}''(\lambda) \cdot (m - s)^{2}}\big),\\
    R^{t}(m) - R^{*}(m)
    & ~=~ m \cdot F_{1}^{*}(m) \cdot (1 - F_{2}^{t}(m)) \cdot \big(1 - e^{-\frac{1}{2}H_{2}^{*}{}''(\mu) \cdot (m - t)^{2}}\big).
\end{align*}

We deduce that $m \cdot F_{1}^{*}(m) \cdot (1 - F_{2}^{t}(m)) \ge \frac{7}{8} \cdot F_{1}^{*}(\frac{7}{8}) \cdot (1 - F_{2}^{t}(\frac{15}{16})) = \frac{7}{8} \cdot F_{1}^{*}(\frac{7}{8}) \cdot (1 - F_{2}^{*}(\frac{15}{16})) \approx 0.1647 > \frac{4}{25}$.
Together with $\min\big(H_{2}^{*}{}''(\lambda),\ H_{2}^{*}{}''(\mu)\big) \le H_{2}^{*}{}''(\frac{7}{8}) \approx 61.4086 > 60$ (\Cref{claim:two-MHR:first-order:1}), we can conclude with
\begin{align*}
    R^{t}(m) - R^{*}(m)
    &\textstyle ~\ge~ \frac{4}{25} \cdot \big(1 - e^{-30 \cdot \max((m - s)^{2},\ (m - t)^{2})}\big)\\
    &\textstyle ~\ge~ \frac{4}{25} \cdot \big(1 - e^{-\frac{15}{2}\eps^{2}}\big)\\
    &\textstyle ~\ge~ \eps^{2}.
\end{align*}
Here the second step uses $(t - s = \eps) \implies \max\big((m - s)^{2},\ (m - t)^{2}\big) \ge \frac{1}{4}\eps^{2}$. And the last step uses $\eps \in (0, \frac{1}{100}) \implies 1 - e^{-\frac{15}{2}\eps^{2}} \ge \frac{25}{4}\eps^{2}$.

The second part ``$R^{t}(x) = R^{*t}(x) \le \frac{3}{4}$ for $x \notin (s, t]$'' is obvious. This finishes the proof of \Cref{lem:two-MHR:revenue}.
\end{proof}

Secondly, the following \Cref{lem:two-MHR:identify} is a counterpart of \Cref{lem:three-MHR:identify}.

\begin{lemma}[Identification Lower Bounds]
\label{lem:two-MHR:identify}
To identify hard instances $\{\bF^{i}\}_{i \in [K]}$ each with probability $\ge \frac{2}{3}$, an identification algorithm $\+B$ makes at least $\bb{E}^{*}[T] = \Omega(\eps^{-5})$ many pricing queries on the base instance $\bF^{*}$ (in expectation over the randomness of both $\+B$ itself and $\bF^{*}$).
\end{lemma}

\begin{proof}
Consider the base instance $\bF^{*}$ and a specific hard instance $\bF^{i}$.
As mentioned, only pricing queries within the corresponding modification interval $(s^{i}, t^{i}]$ can help identify $\bF^{i}$.

Recall that $\DKL(p, q) = p \ln(\frac{p}{q}) + (1 - p) \ln(\frac{1 - p}{1 - q})$ denotes the KL divergence between two Bernoulli distributions with parameters $p, q \in [0, 1]$.
For $x \in (s^{i}, t^{i}] \subseteq [\frac{7}{8}, \frac{15}{16}]$, {we have $0 \le F^{*}(x) - F^{i}(x) \le 20\eps^{2} \le \frac{1}{500}$ (\Cref{lem:two-MHR:first-order} and $\eps \in (0, \frac{1}{100})$)} and $F^{*}(x) = 1 - \frac{3}{4x} \in [\frac{1}{7}, \frac{1}{5}]$, so \Cref{cla:DKL-UB} is applicable and gives
\begin{align*}
    \textstyle
    \DKL\big(F^{*}(x),\ F^{i}(x)\big)
    ~\le~ 3\cdot (20\eps^{2})^{2}
    ~=~ 1200\eps^{4}.
\end{align*}
Then, regarding the event $\@E^{i} \defeq \{\text{$\+B$ outputs $\bF^{i}$}\}$, we know from \Cref{lem:kl-bound} that
\begin{align*}
    \textstyle
    \DKL\big(\bb{P}^{*}[\@E^{i}],\ \bb{P}^{i}[\@E^{i}]\big)
    ~\le~ \DKL\big(F^{*}(x),\ F^{i}(x)\big) \cdot \bb{E}^{*}[T^{i}]
    ~\le~ 1200\eps^{4} \cdot \bb{E}^{*}[T^{i}].
\end{align*}
By enumerating all $i \in [K]$, we can upper-bound the sum $\sum_{i \in [K]} \DKL(\bb{P}^{*}[\@E^{i}],\ \bb{P}^{i}[\@E^{i}])$ as follows:
\begin{align*}
    \textstyle
    \sum_{i \in [K]} \DKL\big(\bb{P}^{*}[\@E^{i}],\ \bb{P}^{i}[\@E^{i}]\big)
    ~\le~ {\sum_{i \in [K]} 1200\eps^{4} \cdot \bb{E}^{*}[T^{i}]
    ~\le~ 1200\eps^{4} \cdot \bb{E}^{*}[T].}
\end{align*}
Here the last step uses the linearity of expectations and that $\sum_{i \in [K]} T_{i} \le T$ (almost surely over all possible randomness).

Moreover, since the KL divergence $\DKL(p, q)$ is a convex function (\Cref{cla:DKL-convex}), using Jensen's inequality (\Cref{cla:Jensen}), we can lower-bound the sum $\sum_{i \in [K]} \DKL(\bb{P}^{*}[\@E^{i}],\ \bb{P}^{i}[\@E^{i}])$ as follows:
\begin{align*}
    \textstyle
    \sum_{i \in [K]} \DKL\big(\bb{P}^{*}[\@E^{i}],\ \bb{P}^{i}[\@E^{i}]\big)
    &\textstyle ~\ge~ K \cdot \DKL\Big(\frac{\sum_{i \in [K]} \bb{P}^{*}[\@E^{i}]}{K},\ \frac{\sum_{i \in [K]} \bb{P}^{i}[\@E^{i}]}{K}\Big)\\
    &\textstyle ~\ge~ K \cdot \DKL(\frac{1}{6}, \frac{2}{3})\\
    &\textstyle ~\ge~ \frac{1}{2}K.
\end{align*}
Here the second step uses ``$\{\@E^{i}\}_{i \in [K]}$ are disjoint'' $\implies \frac{\sum_{i \in [K]} \bb{P}^{*}[\@E^{i}]}{K} \le \frac{1}{K}\le \frac{1}{6}$ and the premise of the lemma ``$\bb{P}^{i}[\@E^{i}] \ge \frac{2}{3}$ for $i \in [K]$''.
And the last step uses $\DKL(\frac{1}{6}, \frac{2}{3}) = \frac{5}{6}\ln 5 - \frac{7}{6}\ln 2 \approx 0.5325$.

Combining the above two equations directly gives $\bb{E}^{*}[T] \ge \frac{1}{2400}\eps^{-4} \cdot K \ge \frac{1}{48000}\eps^{-5}$, where the last step uses $\eps \in (0, \frac{1}{100}) \implies K = \lfloor \frac{1}{16}\eps^{-1} \rfloor \ge \frac{1}{20}\eps^{-1}$. This finishes the proof of \Cref{lem:two-MHR:identify}.
\end{proof}

\begin{proof}[Proof of \Cref{thm:two-MHR}]
Incorporate the ``pricing-to-identification'' reduction into \Cref{lem:two-MHR:identify}: if a pricing algorithm $\+A$ always outputs an $\eps' \defeq \eps^{2} \in (0, \frac{1}{10000})$ approximately optimal price $p^{\+A}$ with probability $\ge \frac{2}{3}$, then it makes at least $\Omega(\eps^{-5}) = \Omega(\eps'{}^{-5 / 2})$ many pricing queries on the base instance $\bF^{*}$.

This finishes the proof of \Cref{thm:two-MHR}.
\end{proof}

\newcommand{\regret}{{\sf Regret}}

\section{Regret Lower Bounds}
\label{sec:regret}

In this appendix, we depart from the pricing query complexity problem and study the \textit{regret minimization} problem.
Recall the \textit{first-order value distribution} $F$ and the \textit{revenue function} $R$; as before, we consider $[0, 1]$-supported value distributions $\bF$, so the \textit{optimal uniform price} $p^{\sf opt} = p^{\sf opt}(\bF) = \argmax_{p \in [0,1]} R(p)$ is well-defined and lies in the support $[0, 1]$.
\begin{align*}
    F(p)
    &\textstyle ~=~ \bb{P}_{\bv \sim \bF}[(\max_{i \in [n]} v_{i}) < p]
    ~=~ \prod_{i = 1}^{n} F_{i}(p),
    &&\textstyle \forall p \ge 0,\\
    R(p)
    &\textstyle ~=~ p \cdot \bb{P}_{\bv \sim \bF}[(\max_{i \in [n]} v_{i}) \ge p]
    ~=~ p \cdot \big(1 - F(p)\big),
    &&\textstyle \forall p \ge 0.
\end{align*}
In the regret minimization problem, an algorithm $\+A$ needs to play a $T$-round repeated game, as follows:
\begin{itemize}
    \item At the beginning, $\+A$ has no information of the value distributions $\bF$ (except for their independence and $[0, 1]$ support).
    
    \item Each round $t = 1, 2, \dots, T$ refers to an \textit{independent} trial $\bv^{t} \sim \bF$ of the {\UniformPricing} mechanism:\\
    $\+A$ posts a price $p^{t}$, acquires whether the sale succeeds or not $z^{t} = \bb{1}[(\max_{i \in [n]} v_{i}^{t}) \ge p^{t}] \in \{0, 1\}$, and thus accumulates an amount of $p^{t} \cdot z^{t}$ revenue.
    
    

\end{itemize}
The regret minimization problem asks for the \textit{minimax regret} $\regret(T) \in [0, T]$ accumulated throughout the game, against the optimal {\UniformPricing} revenue $R(p^{\sf opt})$:
\begin{align*}
    \textstyle
    \regret(T) ~\defeq~ \min_{\+A}\/ \max_{\bF}\/ \bb{E}_{\bF, \+A}\big[\sum_{t = 1}^{T} \big(R(p^{\sf opt}) - p^{t} \cdot z^{t}\big)\big].
\end{align*}

Below in \Cref{subsec:regret:reduction}, we will give a black-box reduction from \textit{pricing query complexity} lower bounds to \textit{regret} lower bounds. As direct consequences, this reduction in combination with \Cref{lem:two-regular:identify,lem:three-MHR:identify,lem:two-MHR:identify} (after suitable scales of $\eps \in (0, 1)$) gives the following \Cref{cor:two-regular,cor:three-MHR,cor:two-MHR}.

\begin{corollary}
\label{cor:two-regular}
For two (or more) regular distributions, the minimax regret of {\UniformPricing} is $\Omega(T^{2 / 3})$.
\end{corollary}

\begin{corollary}
\label{cor:three-MHR}
For three (or more) MHR distributions, the minimax regret of {\UniformPricing} is $\Omega(T^{2 / 3})$.
\end{corollary}

\begin{corollary}
\label{cor:two-MHR}
For two MHR distributions, the minimax regret of {\UniformPricing} is $\Omega(T^{3 / 5})$.
\end{corollary}

\subsection{A Reduction from Identification to Regret Minimization}
\label{subsec:regret:reduction}



Given a sufficiently small $\eps \in (0,1)$ and a sufficiently large $K \defeq K(\eps)$. Suppose that we have one base instance $\bF^{*}$ and $K$ hard instances $\{\bF^{i}\}_{i \in [K]}$ that, for some disjoint intervals $(s^{i}, t^{i})$ for $i\in [K]$, satisfy the following three conditions:
\begin{itemize}
    \item $R^*(p)$ is maximized everywhere over the interval $p \in (s^{1}, t^{K}]$.
    
    \item $R^{*}(p) = R^{i}(p)$ for $p \notin (s^{i}, t^{i})$.
    
    \item $\max_{p \in (s^{i}, t^{i}]} \big(R^{i}(p) - R^{*}(p)\big) \geq \eps$, for each $i \in [K]$.
\end{itemize}
Note that, after suitable scales of $\eps \in (0, 1)$, these conditions hold for each of the lower-bound constructions in \Cref{sec:regular,sec:three-MHR,sec:two-MHR}.

\begin{lemma}[Identification Upper Bounds]
\label{lem:query2regret}
Given a universal constant $\alpha \in (0, 1)$, if there exists an $O(T^{\alpha})$-regret algorithm $\+A$, then there exists an $O(\eps^{-\frac{1}{1 - \alpha}})$-query identification algorithm $\+B$.
\end{lemma}

\begin{proof}
Without loss of generality, for some universal constant $c > 0$, the algorithm $\+A$ has a regret bound $\regret(T) \le c \cdot T^{\alpha}$; we would run it for $T = \lceil (3c / \eps)^{\frac{1}{1-\alpha}} \rceil = O(\eps^{-\frac{1}{1 - \alpha}})$ rounds, and let $T^{i}$ denote how many pricing queries are made within the index-$i$ modification interval $(s^{i}, t^{i})$, for $i \in [K]$.

Denote by $\bb{E}^{i}[\cdot]$ the expectations in each possibility $i \in [K]$. We can deduce that
\begin{align*}
    \textstyle
    \frac{1}{T} \cdot \bb{E}^{i}[T - T^{i}]
    ~\le~ \frac{1}{T} \cdot \eps^{-1} \cdot \regret(T)
    ~\le~ c \cdot \eps^{-1} \cdot T^{\alpha - 1}
    ~\le~ \frac{1}{3}.
\end{align*}
Here the first step uses $\regret(T) \le \eps \cdot \bb{E}^{i}[T - T^{i}]$ (obvious).
The second step uses $\regret(T) \le c \cdot T^{\alpha}$.
And the last step uses $T = \lceil (3c / \eps)^{\frac{1}{1-\alpha}} \rceil$.

As a consequence, the identification algorithm $\+B$ can, after the $T = O(\eps^{-\frac{1}{1 - \alpha}})$ many pricing queries, simply output each $\bF^{i}$ for $i \in [K]$ with probability $T^{i} / T$;\footnote{If there are pricing queries made outside all modification intervals $(s^{i}, t^{i})$ for $i \in [K]$, then $\+B$ can output arbitrarily with the remaining probability $(1 - \sum_{i \in [K]} T_{i} / T)$.} the above equation immediately implies that $\+B$ successes with probability $\ge \frac{2}{3}$. This finishes the proof.
\end{proof}


The following \Cref{cor:query2regret} is a direct implication of \Cref{lem:query2regret}.

\begin{corollary}[Regret Lower Bounds]
\label{cor:query2regret}
Given a universal constant $\beta = \frac{1}{1 - \alpha} > 1$, if any identification algorithm $\+B$ has query complexity $\Omega(\eps^{-\beta})$, then any regret minimization algorithm $\+A$ has regret $\Omega(T^{1 - 1 / \beta})$.
\end{corollary}

\end{document}